\documentclass[letterpaper,twocolumn,10pt]{article}
\usepackage{usenix}
\usepackage{amsmath,amssymb}
\usepackage{graphicx}
\usepackage[table]{xcolor}
\usepackage{xspace}
\usepackage[normalem]{ulem}
\usepackage{booktabs,multirow,makecell,adjustbox,array}
\usepackage{tikz}
\usetikzlibrary{arrows.meta,positioning}
\microtypecontext{spacing=nonfrench}

\newcommand\attackname{\textsc{SpectralLeak}\xspace}
\newcommand\blockhlpname{\textsc{LatticeLeak}\xspace}
\newcommand\sysname{\textsc{ArrowCloak}\xspace}
\newcommand\defensename{\textsc{ButterflyCloak}\xspace}
\newcolumntype{C}[1]{>{\centering\arraybackslash}p{#1}}

\begin{document}

\date{}
% \author{Anonymous Submission}
\title{\Large \bf Beyond Vector Hiding: Breaking and Mitigating Shared-Direction Weight Obfuscation in TEE-Offloaded Large Language Models}

\author{
Menghui Zhang$^{1}$$^{\ast}$,
Aoying Zheng$^{1}$$^{\ast}$,
Guoxiao Liu$^{2}$,
Zizhuang Deng$^{1,3}$,\\
Jiejing Wen$^{1}$$^{\dagger}$,
Jincheng Zhuang$^{1}$$^{\dagger}$,
and Ran Tao$^{1}$\\[5pt]
{$^{1}$School of Cyber Science and Technology,
Shandong University, Qingdao, China}\\
{$^{2}$Tsinghua University, Beijing, China}\\
{$^{3}$Suzhou Research Institute of Shandong University,
Suzhou, China}
}

\maketitle
\begingroup
\renewcommand{\thefootnote}{}
\footnotetext{
$^{\ast}$Equal contribution.
\quad
$^{\dagger}$Corresponding authors.
}
\endgroup

\begin{abstract}
Trusted Execution Environment (TEE)-shielded partitioning of Large Language Models (LLMs) accelerates on-device inference by offloading obfuscated linear layers to an untrusted accelerator while retaining only a small correction inside the TEE. 
However, earlier lightweight obfuscation schemes preserved weight-vector directions and were broken by ArrowMatch.
To defend against this attack, ArrowCloak injects scalar multiples of the
same hidden direction into all weight vectors, enabling lightweight trusted
correction.

We show that this reuse leaves a rank-one relation across the complete accelerator-visible matrix. For the released real-valued scheme, we propose \attackname, which estimates and removes the shared component. Across 12 task settings, its surrogates achieve $87.98\%$ mean accuracy versus $89.85\%$ for the victims. 
In our defense-favorable mod-$Q$ realization of ArrowCloak's published
modular security formulation, mod-$Q$ arithmetic suppresses this spectral
signal but retains the algebraic rank-one relation modulo $Q$.
We therefore propose \blockhlpname, which exploits the resulting hidden lattice. In our BERT-Base and GPT2-Base experiments, it reconstructs every protected fixed-point parameter exactly; across all evaluated architectures, the reconstructed models retain victim-level task accuracy without victim queries, labels, or fine-tuning.

These findings identify \emph{shared rank-one reuse as the root cause of the
leakage exploited by our attacks}.
Guided by this insight, we design \defensename, a keyed maximal-rank butterfly mask that replaces the reused direction with distinct mask rows while retaining fast trusted correction. In our full-BERT evaluation, protection leaves task accuracy unchanged ($91.6284\%$ before and after), reduces ArrowMatch recovery to near-random levels, and prevents \blockhlpname. For a square $n\times n$ layer with $\ell$ input activations, its trusted correction costs $O(n\ell\log n)$, only a logarithmic factor above ArrowCloak's $O(n\ell)$ correction and asymptotically below dense masking's $O(n^2\ell)$ cost.
\end{abstract}

\section{Introduction}
As Large Language Models (LLMs) are increasingly deployed on smartphones, personal computers, and other edge devices, protecting proprietary model weights has become an important security problem. 
Model weights embody substantial investments in data collection, training, and task-specific adaptation, and therefore constitute valuable intellectual property. 
In cloud deployment, users typically have only black-box access through an input--output interface. 
In on-device deployment, however, the model runs on a user-controlled device, potentially exposing its files, memory, and execution process. Prior studies have shown that model weights can be extracted from on-device applications, device memory, or data transfers~\cite{sun2021mindweights,nayan2024sok,rakin2022deepsteal,zhu2021hermes,yuan2024hypertheft}.
Related attacks recover model architectures or protected inputs through shared resources and ciphertext side channels~\cite{yan2020cachetelepathy,yuan2025ciphersteal}, while black-box and transfer-learning attacks demonstrate that public interfaces can also enable functional or parameter recovery~\cite{tramer2016stealing,orekondy2019knockoff,wang2018transfer,chen2022teacher,horwitz2024recovering}.
On-device model protection therefore aims to prevent restricted black-box access from becoming direct white-box access to proprietary weights.
Trusted Execution Environments (TEEs) can isolate model weights and sensitive computation from the rest of the device. 
However, their limited memory and computation make full-model execution expensive. 
TEE-shielded model partitioning addresses this limitation by offloading computation-intensive linear layers to an untrusted accelerator while retaining only lightweight correction inside the TEE~\cite{tramer2019slalom,mo2020darknetz,zhang2024privacy,li2025teeslice}.
Research on GPU and accelerator TEEs seeks to enlarge the trusted computing base~\cite{volos2018graviton,wang2024cage,sridhara2024acai}; nevertheless, partitioned execution remains attractive when the accelerator is outside the trusted boundary or when trusting the complete accelerator is too costly.
Because the accelerator must receive the weights required for computation, these systems obfuscate the weights before offloading them and recover the correct outputs inside the TEE. 
The central challenge is to hide the original weights while keeping the in-TEE output correction lightweight, thereby preserving the efficiency benefits of computation offloading.

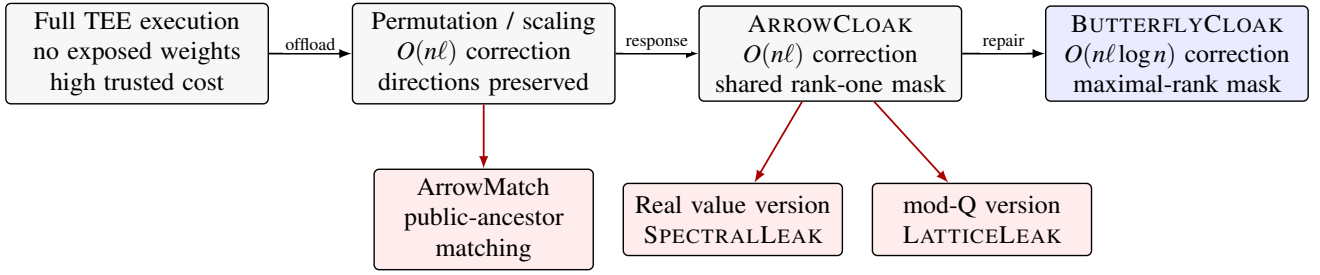
\begin{figure*}[t]
\centering
\resizebox{0.97\textwidth}{!}{%
\begin{tikzpicture}[
  >=latex,
  stage/.style={draw, rounded corners=2pt, align=center, minimum height=13mm,
    text width=34mm, line width=0.55pt, fill=black!3},
  leak/.style={draw, rounded corners=2pt, align=center, minimum height=10mm,
    text width=28mm, line width=0.55pt, fill=red!7},
  repair/.style={stage, fill=blue!8},
  flow/.style={->, line width=0.7pt},
  attack/.style={->, line width=0.7pt, draw=red!65!black},
  edgeword/.style={font=\scriptsize, fill=white, inner sep=1pt, align=center}
]
\node[stage] (tee) at (0,0) {Full TEE execution\\no exposed weights\\high trusted cost};
\node[stage] (pd) at (4.8,0) {Permutation / scaling\\$O(n\ell)$ correction\\directions preserved};
\node[stage] (ac) at (9.6,0) {\sysname\\$O(n\ell)$ correction\\shared rank-one mask};
\node[repair] (bm) at (14.4,0) {\defensename\\$O(n\ell\log n)$ correction\\maximal-rank mask};
\node[leak] (am) at (4.8,-2.3) {ArrowMatch\\public-ancestor matching};
\node[leak] (sl) at (8.25,-2.3) {Real value version\\\attackname};
\node[leak] (bh) at (11.7,-2.3) {mod-Q version\\\blockhlpname};
\draw[flow] (tee) -- node[edgeword,above]{offload} (pd);
\draw[flow] (pd) -- node[edgeword,above]{response} (ac);
\draw[flow] (ac) -- node[edgeword,above]{repair} (bm);
\draw[attack] (pd) -- (am);
\draw[attack] (ac) -- (sl);
\draw[attack] (ac) -- (bh);
\end{tikzpicture}}
\caption{The security--efficiency evolution studied in this paper.  Earlier
lightweight transformations preserve vector directions; \sysname~ blocks
that leakage by reusing one hidden direction, but the reuse creates a global
  rank-one invariant.  Our two attacks expose its real and finite-field views,
  and \defensename~ explores a maximal-rank repair. Cost labels use the
  square-layer shorthand $m=d=n$.}
\label{fig:security_evolution}
\end{figure*}

Existing schemes face a clear tension between security and efficiency. Lightweight permutation, scaling, and channel-wise transformations require only $O(n\ell)$ trusted correction, but preserve weight-vector directions~\cite{li2024translinkguard,shen2022soter,xu2024tempo,sun2023shadownet,sun2025tsqp}.
Active weight-obfuscation defenses can also perturb selected parameters, but must preserve model utility and efficient inference~\cite{zhou2023nnsplitter}.
More expressive schemes such as GroupCover~\cite{zhang2024groupcover} can change vector directions, but incur substantially higher trusted computation.
ArrowMatch~\cite{wang2025game} showed that the former lightweight schemes can be broken by matching the directions of exposed private-model vectors to those of a public pre-trained model. 
Because private models are commonly fine-tuned from public checkpoints, corresponding weights often retain exploitable structure~\cite{wang2018transfer,chen2022teacher,horwitz2024recovering}; the recovered correspondences can therefore be used to construct functional surrogate models.
To defend against ArrowMatch, ArrowCloak~\cite{wang2025game} injects the same hidden direction into every private weight vector using different coefficients, followed by scaling and permutation, and establishes a reduction to Learning With Errors (LWE) problem \cite{regev2005lwe}.
This shared component disrupts per-vector direction matching while retaining $O(n\ell)$ trusted correction. 
When vectors are considered independently, ArrowCloak therefore appears to provide both direction protection and lightweight overhead.
\emph{However, we find that hiding individual weight vector directions does not protect the complete weight matrix.}
Because ArrowCloak injects the same hidden direction into all vectors, this reuse leaves a shared rank-one relation in the complete matrix. 
In row notation, the transformation is $Y=P(D+rk^{\top})W$, where $W$ is the private weight matrix, $D$ is diagonal, $P$ is a permutation, and $rk^{\top}$ captures the shared rank-one reuse. 
This relation remains in both the real-valued implementation of ArrowCloak and its modular variant.

For the original ArrowCloak over the reals, we propose \attackname, which removes the shared component and recovers surrogate models that nearly match victim-model utility. 
Across 12 task settings covering 4 models and 7 datasets, the recovered models achieve $87.98\%$ mean accuracy, compared with $89.85\%$ for the victims. 
A separate scale extension to Qwen3-8B, Llama-3.1-8B, and Gemma-4-E2B obtains $96.22$--$96.56\%$ without public-data refinement, at most $0.34$ percentage points from the corresponding victims.
Although field-wide mod-$Q$ arithmetic defeats \attackname, the underlying rank-one relation remains. 
We therefore propose \blockhlpname for the modular variant. 
In our BERT-Base and GPT2-Base experiments, \blockhlpname reconstructs every protected fixed-point parameter exactly. 
It also restores victim-level accuracy in the evaluated ViT and GPT2-XL settings. 
The algebraic attack requires no victim queries, labels, or fine-tuning.
These findings identify \emph{shared rank-one reuse as the root cause of the leakage}. 
Guided by this insight, we further design \defensename, a keyed maximal-rank butterfly mask that removes rank-one reuse without dense TEE computation. 
It leaves full-BERT accuracy unchanged. Across the three larger models, it suppresses the tested spectral and correspondence attacks and keeps simulated protected accuracy within $0.23$ percentage points of the victim. 
For an $m\times d$ layer, its trusted correction costs $O(\ell\max\{m,d\}\log d)$, compared with ArrowCloak's $O(\ell(m+d))$ and dense masking's $O(\ell md)$.

Figure~\ref{fig:security_evolution} summarizes this evolution: earlier lightweight schemes preserve weight-vector directions; ArrowCloak blocks this directional leakage but introduces shared rank-one reuse; \attackname and \blockhlpname expose this structure over the reals and finite fields, respectively; and \defensename eliminates the root cause of the leakage.

Two recent concurrent preprints also analyze ArrowCloak.
MOSAIC~\cite{chiang2026mosaic} questions ArrowCloak's claimed security
reduction to LWE. After we registered our USENIX Security submission on
August 18, 2026, another concurrent preprint was submitted to arXiv on
August 21, 2026, introducing ArrowRevelio, an SVD-based attack against
ArrowCloak's released real-valued implementation~\cite{liu2026hidingdirectionsleakingstructure}.
Compared with these concurrent works, we study ArrowCloak in both settings:
we propose a PCA attack against its original implementation over the reals
and use lattice reduction to attack its modular formulation. In addition, we design a mitigation against these attacks.

Our contributions are summarized as follows:
\begin{itemize}
  \item \textbf{Matrix-level leakage.}
  We identify shared rank-one reuse as the root cause of information leakage in both ArrowCloak and its modular variant, showing that security must be analyzed over the complete accelerator-visible matrix rather than individual vectors.
  \item \textbf{Attacks across arithmetic settings.}
  We propose \attackname for the real-valued scheme, recovering surrogates with $87.98\%$ mean accuracy versus $89.85\%$ for the victims across 12 task settings and extending to three 5.1--8.2B-parameter models without refinement. We further propose \blockhlpname for the modular variant; it exactly reconstructs all protected fixed-point parameters for BERT-Base and GPT2-Base and restores victim-level utility across the evaluated BERT, ViT, GPT2-Base, and GPT2-XL settings without victim queries, labels, or fine-tuning.
  \item \textbf{Mitigation of rank-one reuse.}
  We design \defensename, a keyed maximal-rank butterfly mask that eliminates shared rank-one reuse. On full BERT it suppresses all evaluated attacks, including \blockhlpname. On three larger language models it preserves utility and suppresses the tested spectral and correspondence attacks, with $O(\ell\max\{m,d\}\log d)$ trusted correction.
\end{itemize}

\section{Background and Preliminaries}
\label{sec:background}

\subsection{TEE-Shielded Model Partitioning}
In on-device deployment, a Trusted Execution Environment (TEE) isolates model
weights and sensitive computation from untrusted software. However, existing TEEs typically have limited computational capacity and available memory, making it difficult to efficiently execute the large-scale matrix operations required by modern models. To reduce the overhead of full-model execution inside the TEE, prior work has proposed TEE-shielded model partitioning, which offloads computation-intensive linear layers to a GPU or another hardware accelerator outside the TEE while retaining sensitive state and necessary output correction within the TEE. Slalom masks outsourced linear computation and verifies its execution using Freivalds' algorithm~\cite{tramer2019slalom}; DarkneTZ partitions on-device models between trusted and untrusted execution~\cite{mo2020darknetz}; and subsequent work further analyzes the potential model information leakage introduced by such partitioning~\cite{zhang2024privacy}.

\begin{figure}[t]
  \centering
  \includegraphics[width=\columnwidth]{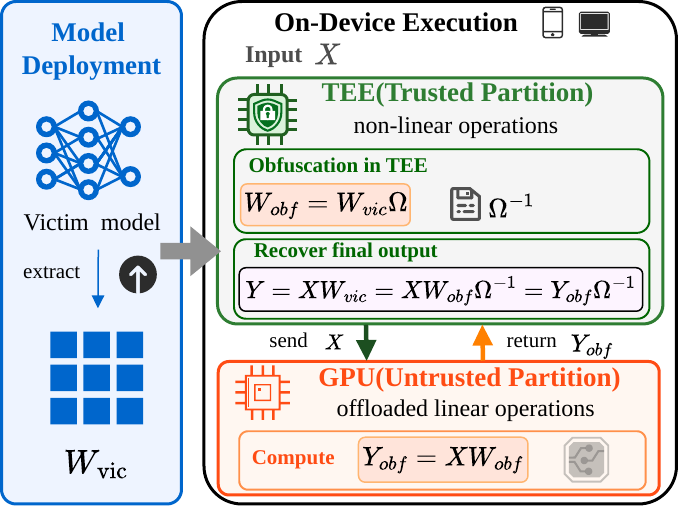}
  \caption{Cooperative execution between the TEE and an untrusted accelerator. The accelerator performs linear computation using obfuscated weights, while the TEE recovers the correct outputs.}
  \label{fig:tee_gpu_execution}
\end{figure}

Figure~\ref{fig:tee_gpu_execution} illustrates cooperative execution between
the TEE and the accelerator. For a private weight matrix $W_{\mathrm{vic}}$ to be offloaded, the system generates an obfuscated matrix $W_{\mathrm{obf}}$ during deployment or initialization and provides it to the untrusted accelerator. During inference, the accelerator computes with $W_{\mathrm{obf}}$ and
returns the result, which the TEE corrects using its retained secret
information. In this way, the computation-intensive matrix multiplication is executed by the accelerator, while the TEE performs only lightweight output correction.
We focus on whether the complete obfuscated weight matrices legitimately provided to the untrusted accelerator reveal the private model. Unlike query-based model extraction, memory attacks, or TEE side channels
\cite{tramer2016stealing,orekondy2019knockoff,rakin2022deepsteal,zhu2021hermes},
our attacks use only these matrices and require no additional leakage. Input protection, result integrity, and adversarial capabilities are specified in our threat model.

\subsection{Weight Obfuscation}
\label{sec:Weight_Obfuscation}

\textbf{Notation.}
Unless otherwise stated, weight matrices are represented in row-oriented
form throughout this paper. For
$W,Y,U\in\mathcal{D}^{m\times d}$, each row
$w_i,y_i,u_i\in\mathcal{D}^{1\times d}$ denotes one weight vector, where
$\mathcal{D}=\mathbb{R}$ for the released implementation and
$\mathcal{D}=\mathbb{F}_Q$ for the modular realization.

Throughout the matrix-level analysis, weight vectors are represented as row
vectors, while all other vector-valued quantities are represented as column
vectors unless explicitly stated otherwise. Thus, directions such as
$v$ and $\widehat v$, as well as coefficient vectors such as $r$, $k$, and
the blockwise coefficients introduced later, are column vectors. Their row
forms are written using the transpose operator when needed. Consequently,
weight-row projections act from the right, and for
$X\in\mathcal{D}^{\ell\times d}$ the accelerator computes
$O_{\mathrm{obf}}=XY^{\top}$.

\begin{figure}[t]
  \centering
  \includegraphics[width=\columnwidth]{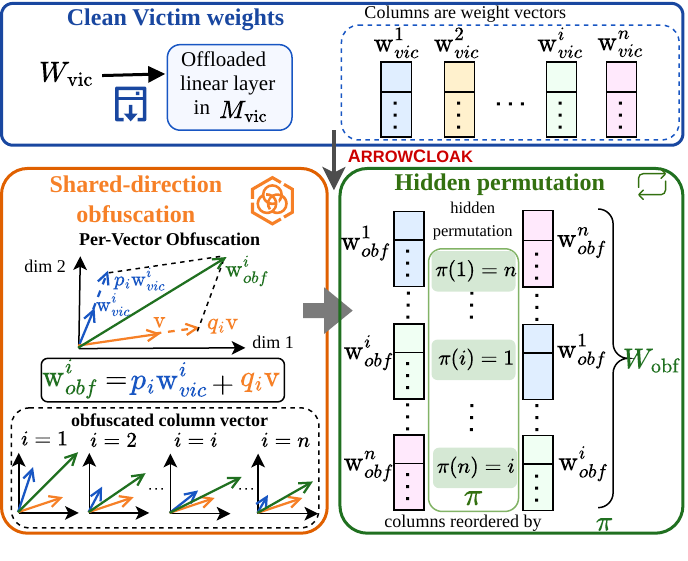}
  \caption{Weight obfuscation in ArrowCloak, shown using the original paper’s column-vector notation. All weight vectors reuse the same hidden direction with different injection coefficients before permutation.}
  \label{fig:arrowcloak_obfuscation}
\end{figure}

Figure~\ref{fig:arrowcloak_obfuscation} and its accompanying per-vector
equation follow the original column-vector notation used by ArrowCloak.
Figure~\ref{fig:tee_gpu_execution} is a schematic system overview using local
system-level notation. All matrix-level analysis thereafter follows the
convention above.

Weight obfuscation must balance security and efficiency. On the one hand, the obfuscated weights should not reveal information about the original model. On the other hand, output correction inside the TEE must remain lightweight; otherwise, computation offloading provides little benefit. For a linear layer of width $n$ and sequence length $\ell$, the original matrix multiplication costs $O(n^2\ell)$. The cost of weight obfuscation and output correction should therefore be substantially lower than this complexity.
Existing lightweight schemes use scaling, permutation, and channel
transformations to retain $O(n\ell)$ trusted correction
\cite{shen2022soter,sun2023shadownet,li2024translinkguard,xu2024tempo,sun2025tsqp}, but preserve exploitable weight-vector directions.
ArrowMatch exploits this information by matching exposed vectors to corresponding vectors in a public model~\cite{wang2025game}. Because private models are commonly fine-tuned from public models, their corresponding weight vectors often remain close in direction. ArrowMatch uses this similarity to recover vector correspondences and construct a functional surrogate model.
GroupCover instead changes vector directions through dense mixing within groups, but increases trusted correction to $O(n^2\ell/k)$~\cite{zhang2024groupcover}. Although it was not evaluated as an ArrowMatch target, GroupCover illustrates the higher trusted cost required by dense direction mixing.
Table~\ref{tab:design_landscape} summarizes the relevant tradeoffs among representative weight obfuscation designs.
\begin{table}[t]
\centering
\caption{Representative weight obfuscation designs. Costs denote trusted output correction for a layer of width $n$ and sequence length $\ell$.}
\label{tab:design_landscape}
\begingroup
\footnotesize
\setlength{\tabcolsep}{3.5pt}
\resizebox{\columnwidth}{!}{%
\begin{tabular}{@{}l|l|l|l@{}}
\toprule
\textbf{Design} & \textbf{Mixing} & \textbf{TEE work} &
\textbf{Relevant structure} \\
\midrule
Permutation / scaling
& Per vector
& $O(n\ell)$
& Directions preserved \\

GroupCover
& Dense group mixing
& $O(n^2\ell/k)$
& Directions changed \\

\sysname
& Shared direction
& $O(n\ell)$
& Rank-one reuse \\
\bottomrule
\end{tabular}}
\endgroup
\end{table}

To defend against direction matching, Wang et al. proposed ArrowCloak, which injects the same hidden direction into every weight vector while retaining lightweight correction~\cite{wang2025game}. To remain consistent with the presentation in the original ArrowCloak paper,
Figure~\ref{fig:arrowcloak_obfuscation} and Equation~\eqref{eq:arrowcloak_vector} temporarily use its
original column-vector notation. In this notation, the $i$-th private weight
vector $w_i$ is transformed as
\begin{equation}
y_i=p_iw_i+q_iv,
\qquad
v=\sum_j k_jw_j,
\label{eq:arrowcloak_vector}
\end{equation}
where $p_i$ scales the original vector, $q_i$ controls the injection of $v$, and $k_j$ are the coefficients used to construct the hidden direction. ArrowCloak then secretly permutes the transformed vectors.
As shown in Figure~\ref{fig:arrowcloak_obfuscation}, the injection changes
each vector's direction and disrupts its correspondence with the public model.
Because all vectors reuse the same $v$, the TEE computes $Xv$ only once,
keeping the correction cost at $O(n\ell)$.

After this local presentation of the original ArrowCloak notation, we return
to the row-oriented matrix convention defined above. Let
$W,Y\in\mathcal{D}^{m\times d}$ denote the private and obfuscated weight
matrices, respectively. The domain $\mathcal{D}$ is $\mathbb{R}$ for the
released implementation and $\mathbb{F}_Q$ for the modular realization.

For an input activation matrix
$X\in\mathcal{D}^{\ell\times d}$, the accelerator computes
\begin{equation}
    O_{\mathrm{obf}}
    =
    XY^{\top}.
    \label{eq:obfuscated_output}
\end{equation}

The released ArrowCloak implementation operates over the reals. However, the original ArrowCloak paper explicitly quantizes its security relation over a large prime field and formulates the resulting relation in an LWE-style finite-field form \cite{wang2025game}. \blockhlpname is our attack, but the modular target it analyzes is derived from this original security formulation rather than introduced as a new defense by us. Because \cite{wang2025game} does not specify a complete end-to-end modular inference protocol, we instantiate the omitted fixed-point and field-arithmetic details in a defense-favorable manner.

\subsection{Preliminaries for Matrix-Level Analysis}

Our analysis examines the matrix-level structure induced by ArrowCloak's
reuse of a shared hidden direction. We formalize this structure in
Section~\ref{sec:rank_one_structure}. Over the reals, we use principal
component analysis to characterize its dominant spectral manifestation.
Over a finite field, we use code-lattice constructions and lattice basis
reduction to analyze linear relations that remain under modular arithmetic.
We briefly introduce the tools needed for both analyses.

\noindent\textbf{Principal Component Analysis over the Reals.}
For a centered matrix, the first principal component is the leading right
singular vector and captures the direction of maximum projected variance
\cite{jolliffe2002pca,jolliffe2016pca}. We use it to identify a component
shared across the accelerator visible weight rows. Appendix~\ref{sec:appendix_spectral} provides the
formal definition and detailed spectral diagnostics.

\noindent\textbf{Code-Lattice Construction over Finite Fields.}
Let $Q$ be a prime number. Mod-$Q$ arithmetic disrupts the numerical ordering present over the reals, preventing the direct application of real-valued spectral analysis. To analyze the linear relations preserved over a finite field, we use Construction A to embed a linear code into an integer lattice. For a linear code $\mathcal{C}\subseteq\mathbb{F}_Q^d$, the corresponding Construction A lattice is defined as
\begin{equation}
\Lambda(\mathcal{C})
=
\left\{
z\in\mathbb{Z}^d:
z\bmod Q\in\mathcal{C}
\right\}.
\label{eq:construction_a}
\end{equation}
This is the standard Construction A code-to-lattice embedding~\cite{conway1999sphere}.
The LLL algorithm finds short vectors in a given lattice basis~\cite{lenstra1982factoring} and has been widely used to analyze integer structures hidden within modular relations~\cite{nguyen2001twofaces}. We describe how \blockhlpname uses these tools to recover hidden row spaces in Section~ \ref{sec:blockhlp}.

\section{Security Goal and Threat Model}
\label{sec:threat_model}
We follow the practical setting of TEE-shielded model partitioning, where an untrusted accelerator receives obfuscated weights for linear computation, while secret parameters and sensitive operations remain inside the TEE
\cite{tramer2019slalom,mo2020darknetz,zhang2024privacy,wang2025game}.

\noindent\textbf{Scenario.}
A model owner fine-tunes a public pretrained model $M_{\mathrm{pre}}$ into a
private victim model $M_{\mathrm{vic}}$ and deploys it using the partition in
Figure~\ref{fig:tee_gpu_execution}. An untrusted accelerator executes
computation-intensive linear layers using obfuscated weights, while secret
state and lightweight output correction remain inside the TEE. The protected
assets are the private weights and the model functionality they encode.

\noindent\textbf{Adversary's Goal.}
The adversary aims to recover a useful substitute for $M_{\mathrm{vic}}$ rather than every secret obfuscation parameter. We consider two forms of success. \emph{Functional recovery} means that a reconstructed surrogate achieves task accuracy close to that of the victim model. \emph{Exact recovery} means that the recovered weights equal the defender's fixed point representation element by element. Recovering the secret obfuscation parameters is not required under either criterion.

\noindent\textbf{Adversary's Capability.}
The adversary controls the host and untrusted accelerator, records every
obfuscated matrix exposed outside the TEE, and knows the architecture, tensor
shapes, public parameters, and public ancestor $M_{\mathrm{pre}}$, consistent
with ArrowCloak~\cite{wang2025game}. \attackname may use at most $1\%$ of the
public task training split for fine-tuning, whereas \blockhlpname requires no
victim queries, labels, or fine-tuning.
The adversary knows neither the private weights nor the secret permutation,
scales, injection coefficients, hidden direction, or generation seed, and has
no victim prediction oracle. Ground-truth weights and secret parameters are
used only for evaluation, never as attack inputs.

\noindent\textbf{Defender's Goal.}
The defender aims to prevent the exposed matrices from revealing private
weights or enabling reconstruction of a functional surrogate. The protection
must also preserve correct inference and keep trusted correction substantially
below the $O(n^2\ell)$ cost of the original linear computation; otherwise,
offloading provides little benefit. The goal is therefore to protect model
confidentiality without moving dense matrix computation back into the TEE.

\noindent\textbf{Defender's Capability.}
The defender controls model provisioning and the TEE. Before offloading, it transforms the private weights using ArrowCloak or our proposed \defensename and keeps all secret transformation parameters inside the TEE. We assume that the transformation design, model architecture, and public system parameters are known to the adversary, while the secret coefficients, permutations, and keyed parameters remain protected.

\noindent\textbf{Scope.}
We evaluate both the released ArrowCloak implementation over the reals and a field-wide mod-$Q$ realization constructed according to its published LWE-based security analysis. The modular realization is constructed by this work and is not part of the released implementation. Our attacks apply to shared direction schemes whose exposed matrices have the form
$Y=P(D+rk^{\top})W$; we do not claim that all weight obfuscation schemes or all TEE partitioning systems are insecure.
We consider only model confidentiality and assume no TEE compromise, physical
attack, trusted-memory leakage, rollback, or malformed inference. Input and
output privacy, integrity, availability, and generic attacks against the
underlying TEE are out of scope. Our evaluation of \defensename demonstrates
resistance to the concrete attacks studied here, not a general cryptographic
security guarantee.

\section{Shared Rank-One Reuse}
\label{sec:rank_one_reuse}

\subsection{A Diagonal-Plus-Rank-One Structure}
\label{sec:rank_one_structure}

Given the notation defined in Section~\ref{sec:Weight_Obfuscation}, ArrowCloak samples an invertible
diagonal matrix $D$, coefficient vectors $r$ and $k$, and a row permutation
$P$. The shared hidden direction is determined by
\begin{equation}
    v^{\top}=k^{\top}W.
    \label{eq:shared_direction}
\end{equation}
The exposed matrix can therefore be expressed as
\begin{equation}
    Y
    =
    P\bigl(DW+rv^{\top}\bigr)
    =
    P(D+rk^{\top})W.
    \label{eq:arrowcloak_matrix}
\end{equation}

Because $D$ is invertible, $D+rk^{\top}$ is a rank-one update of a diagonal
matrix. Whenever $1+k^{\top}D^{-1}r\neq0$, its inverse follows the standard
Sherman--Morrison identity~\cite{sherman1950adjustment}; ArrowCloak rejects
the singular parameter case so that trusted correction remains well defined.

The injected component $rv^{\top}$ has rank at most one. Although each row
uses a different scaling and injection coefficient, every exposed row
contains a scalar multiple of the same row-form direction $v^{\top}$.
The secret row permutation does not remove this shared low-rank dependence. Thus, although the transformation perturbs each vector individually, the complete exposed matrix remains coupled through the same rank-one component.

This reuse is also what makes correction lightweight. Because every injected
row depends on the same $v$, the TEE computes the corresponding correction
once and reuses it across all rows, retaining $O(n\ell)$ trusted work.
Thus, lightweight correction and shared rank-one reuse stem from the same
design choice. The central security question is whether this matrix-level
relation remains observable under the arithmetic exposed to the attacker.

\subsection{Real Valued Leakage View}
\label{sec:real_leakage_view}
In the released implementation over the reals, each injected row component
has the form $r_i v^{\top}$. Its magnitude varies with $r_i$, but its
direction is always $v$. When this shared variation dominates the row-specific
variation in $DW$, the first principal component of the centered exposed
matrix estimates the reused direction. This creates a direct spectral
manifestation of shared rank-one reuse under the released parameter regime.

Estimating the shared direction is useful because removing it exposes the
private component hidden by the injection. Let $\hat v$ denote the PC1
estimate of $v$, and define the orthogonal projector
\begin{equation*}
\Pi_{\perp\hat v}
=
I-\frac{\hat v\hat v^{\top}}{\|\hat v\|_2^2}.
\end{equation*}
Before the secret row permutation, an exposed row is
$y_i=d_iw_i+r_i v^{\top}$. Because $w_i$ and $y_i$ are row vectors, projection
is applied on the right:
\begin{equation*}
y_i\Pi_{\perp\hat v}
=
d_iw_i\Pi_{\perp\hat v}
+
r_i v^{\top}\Pi_{\perp\hat v}
\approx
d_iw_i\Pi_{\perp\hat v}.
\end{equation*}
The approximation holds when $\hat v$ accurately estimates the reused
direction. The remaining factor $d_i$ changes magnitude but not direction.
Consequently, removing the shared component restores the directional relation
between private rows and their counterparts in the public ancestor.

We verify that this restoration is specific to the leaked shared component,
rather than a generic consequence of dimensionality reduction. For matching
exposed and public rows, we compare cosine distance in the original space
($\delta_{\mathrm{ori}}$), after removing a random direction
($\delta_{\mathrm{rand}}$), and after removing PC1
($\delta_{\mathrm{pc1}}$). Lower distance indicates stronger directional
alignment.

\begin{table}[t]
\centering
\caption{Restoration of direction similarity after PC1 removal.}
\label{tab:projection_restores_similarity}
\begingroup
\footnotesize
\renewcommand{\arraystretch}{1.08}
\setlength{\tabcolsep}{3.5pt}
\begin{tabular}{@{}C{0.18\columnwidth}|C{0.13\columnwidth}|C{0.15\columnwidth}|C{0.16\columnwidth}|C{0.15\columnwidth}@{}}
\specialrule{1.2pt}{0pt}{2pt}
\textbf{Model} & \textbf{Dataset}
& $\boldsymbol{\delta}_{\mathrm{ori}}\downarrow$
& $\boldsymbol{\delta}_{\mathrm{rand}}\downarrow$
& $\boldsymbol{\delta}_{\mathrm{pc1}}\downarrow$ \\
\midrule

\multirow{3}{*}{\textbf{ViT-Base}}
& C10  & 0.8992 & 0.8992 & \textbf{0.0011} \\
& C100 & 0.8992 & 0.8992 & \textbf{0.0013} \\
& F101 & 0.8992 & 0.8992 & \textbf{0.0016} \\
\hline

\multirow{4}{*}{\textbf{BERT-Base}}
& MNLI & 0.8993 & 0.8993 & \textbf{0.0022} \\
& QQP  & 0.8994 & 0.8994 & \textbf{0.0033} \\
& SST2 & 0.8992 & 0.8992 & \textbf{0.0011} \\
& QNLI & 0.8992 & 0.8992 & \textbf{0.0016} \\
\hline

\multirow{4}{*}{\textbf{GPT2-Base}}
& MNLI & 0.8998 & 0.8998 & \textbf{0.0017} \\
& QQP  & 0.8997 & 0.8998 & \textbf{0.0015} \\
& SST2 & 0.8997 & 0.8997 & \textbf{0.0014} \\
& QNLI & 0.8997 & 0.8998 & \textbf{0.0015} \\
\hline

\textbf{GPT2-XL}
& SST2 & 0.9042 & 0.9042 & \textbf{0.0005} \\
\hline

\textbf{Avg.}
& -- & 0.8998 & 0.8998 & \textbf{0.0016} \\
\specialrule{1.2pt}{2pt}{0pt}
\end{tabular}

\vspace{2pt}
\parbox{0.95\columnwidth}{\scriptsize
\textit{Note:}
$\delta_{\mathrm{ori}}$, $\delta_{\mathrm{rand}}$, and
$\delta_{\mathrm{pc1}}$ denote cosine distance in the original space,
after removing a random direction, and after removing PC1, respectively.
Lower is better.
}
\endgroup
\end{table}

Table~\ref{tab:projection_restores_similarity} shows that removing a random
direction leaves the average cosine distance unchanged at $0.8998$, whereas
removing PC1 reduces it to $0.0016$. The same pattern holds across all twelve
task settings. Thus, ArrowCloak does not eliminate the directional relation
between the victim and its public ancestor; in the released implementation,
it largely hides that relation behind one reusable matrix-level component.
Section~\ref{sec:spectralleak} develops this observation into the complete
\attackname attack.

\subsection{Leakage under Mod-$Q$ Arithmetic}
\label{sec:finite_field_leakage_view}

\noindent\textbf{Mod-$Q$ instantiation.}
To determine whether the leakage persists beyond real arithmetic, we further
evaluate the shared direction transformation using field-wide mod-$Q$
arithmetic. Both the encoded weights and the obfuscation coefficients are
computed modulo $Q$.

Let $\widehat W$ denote the original real weight matrix, let $f=16$ denote
the fixed-point exponent, and let $s=2^f$ denote the corresponding
public scale. We encode the matrix as
\begin{equation*}
W=\left\lfloor s\widehat W\right\rceil\bmod Q,
\end{equation*}
where $f=16$, $s=2^{16}$, and $Q=2^{31}-1$. Signed values are decoded by
centered lifting followed by division by $s$. The coefficients $D$, $r$, and $k$ are sampled
directly as field elements without further encoding.

The matrix exposed to the accelerator is
\begin{equation}
Y=P(D+rk^{\top})W\pmod Q.
\label{eq:finite_field_structure}
\end{equation}
We evaluate two coefficient settings. The first maps the small integer ranges
used by the released implementation into $\mathbb F_Q$. The second is a
stronger, defense-favorable setting in which $r$ and $k$ are sampled uniformly
from $\mathbb F_Q$, and the diagonal entries of $D$ are sampled uniformly from
$\mathbb F_Q^{\times}$. We reject samples for which $D+rk^{\top}$ is singular,
ensuring exact inversion inside the TEE.

\noindent\textbf{Spectral suppression.}
Under the field-wide setting, additions and multiplications wrap modulo $Q$
throughout the exposed matrix. Mapping the resulting residues back to real
values does not preserve the Euclidean geometry that makes the shared
direction spectrally dominant. As shown in Section~\ref{sec:evaluation}, this
removes the PC1 signal exploited by \attackname and reduces correspondence
recovery to near-random levels. Field-wide mod-$Q$ arithmetic therefore
defeats the real-valued spectral attack.

Modular wrap removes the real-valued manifestation of rank-one reuse, but not
the underlying relation. Equation~\eqref{eq:finite_field_structure} still
holds algebraically over $\mathbb F_Q$. Thus, although the relation no longer
appears as a dominant principal direction, all exposed rows remain coupled
through the same rank-one structure. PCA can no longer recover the shared
direction, but finite-field linear combinations can still cancel it.

\begin{figure*}[t]
  \centering
  \includegraphics[width=0.96\textwidth]{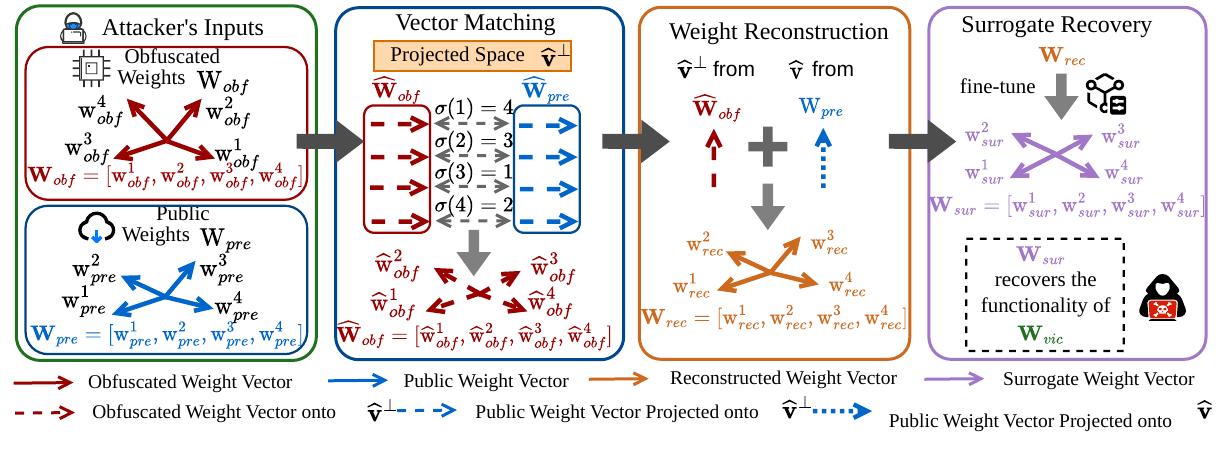}
  \caption{Overview of \attackname against the released implementation over
  the reals. The attacker removes the shared component, recovers vector
  correspondence in the projected space, reconstructs the protected weights,
  and assembles a surrogate model.}
  \label{fig:spectral_overview}
\end{figure*}

To expose this cancellation structure, consider a block $I$ containing $b$
exposed rows. After accounting for the hidden row permutation, the block is
\begin{equation*}
Y_I
=
D_IW_{\pi(I)}
+
r_Iv^{\top}
\pmod Q,
\end{equation*}
where $D_I$ and $r_I$ contain the scales and injection coefficients associated
with the block. Any $c\in\mathbb F_Q^b$ satisfying $c^{\top}r_I=0$ cancels the
shared term:
\begin{equation*}
c^{\top}Y_I
=
c^{\top}D_IW_{\pi(I)}
\pmod Q.
\end{equation*}
Provided that $r_I\neq0$, these coefficient vectors form a subspace of
dimension $b-1$. Thus, although modular arithmetic prevents PCA from
recovering the shared direction, it preserves a structured cancellation
space.

This cancellation space provides the algebraic basis for \blockhlpname.
As detailed in Section~\ref{sec:blockhlp}, the attack recovers hidden row
spans from small blocks and intersects overlapping spans to isolate private
weight rows.

\section{\attackname: Breaking Real Valued ArrowCloak}
\label{sec:spectralleak}

Section~\ref{sec:real_leakage_view} showed that shared rank-one reuse produces
a dominant component in the released implementation over the reals. Removing
this component restores the directional relation between the exposed weights
and their public ancestors. We now turn this leakage into an end-to-end model
reconstruction attack.

\attackname requires only the obfuscated matrices visible to the accelerator
and the corresponding public ancestor model. It does not know the victim
weights, scaling or injection coefficients, shared hidden direction, or
secret permutation. As shown in Figure~\ref{fig:spectral_overview}, the attack
has four stages: shared component removal, correspondence recovery, weight
reconstruction, and surrogate construction.

For one protected layer, let
$W\in\mathbb R^{m\times d}$ denote the private victim matrix,
$Y\in\mathbb R^{m\times d}$ the exposed obfuscated matrix, and
$U\in\mathbb R^{m\times d}$ the corresponding public pre-trained matrix.
We write $w_j$, $y_i$, and $u_j$ for their respective rows.  Because the
secret permutation changes row positions, exposed row $y_i$ need not
correspond to victim row $w_i$; we write $\pi(i)=j$ when $y_i$ originates
from victim row $w_j$.

\subsection{Shared Component Removal}
\label{sec:spectral_remove}

For each exposed matrix $Y\in\mathbb R^{m\times d}$, \attackname centers its
rows and computes the first principal component, denoted by $\widehat v$.
Only the one-dimensional subspace spanned by $\widehat v$ is used. The corresponding orthogonal projector is
\begin{equation}
\Pi_{\perp}
=
I-\frac{\widehat v\widehat v^{\top}}
{\|\widehat v\|_2^2}.
\label{eq:spectral_projection}
\end{equation}

After the secret permutation, exposed row $y_i$ corresponds to victim row
$w_{\pi(i)}$ and can be written as
$y_i=d_iw_{\pi(i)}+r_i v^{\top}$, where $d_i$ and $r_i$ denote its scaling
and injection coefficients.  Because the weights are represented as row
vectors, projection is applied on the right:
\begin{equation}
y_i\Pi_{\perp}
=
d_iw_{\pi(i)}\Pi_{\perp}
+
r_i v^{\top}\Pi_{\perp}
\approx
d_iw_{\pi(i)}\Pi_{\perp}.
\label{eq:spectral_projected_relation}
\end{equation}
The approximation holds because $\widehat v$ estimates the reused direction,
making $v^{\top}\Pi_{\perp}\approx0$. Projection therefore suppresses the
component introduced by ArrowCloak while retaining the private component in
the orthogonal subspace.

\subsection{Correspondence Recovery}
\label{sec:spectral_matching}

After removing the shared component, the attacker must recover the secret row
permutation. \attackname projects both the exposed rows $y_i$ and the
corresponding public ancestor rows $u_j$ using $\Pi_{\perp}$. Following the
direction-based comparison used in ArrowCloak~\cite{wang2025game}, we evaluate
each candidate pair in the projected space using
\begin{equation}
s_{ij}
=
\left|
\cos\!\left(
y_i\Pi_{\perp},
u_j\Pi_{\perp}
\right)
\right|.
\label{eq:spectral_match_score}
\end{equation}

The absolute value makes the score invariant to the sign of the unknown scale
$d_i$. If the released implementation restricts $d_i$ to positive values, the
absolute value can be omitted without changing the assignment.

Rather than matching rows independently, \attackname enforces the one-to-one
structure of the secret permutation and solves
\begin{equation}
\widehat\pi
=
\operatorname*{arg\,max}_{\pi\in S_m}
\sum_{i=1}^{m}s_{i,\pi(i)},
\label{eq:spectral_assignment}
\end{equation}
where $S_m$ is the set of permutations of $m$ rows and
$\widehat\pi(i)=j$ denotes the public/victim row position assigned to exposed
row $y_i$.  The true secret correspondence $\pi$ is unknown to the attacker;
$\widehat\pi$ is its estimate recovered from the projected-space scores. The public model is used
only as a correspondence reference; the victim-specific residual information
remains in the projected exposed rows.

The novelty here is not the directional similarity metric itself, which
follows the comparison used in ArrowCloak, but the space in
which it becomes effective. ArrowMatch performs direction matching directly
in the exposed space, where ArrowCloak suppresses the useful cosine signal.
\attackname first removes the leaked shared component and applies the
directional comparison only after the victim--public relation reappears in
the projected space.

\subsection{Weight Reconstruction}
\label{sec:spectral_reconstruction}

\begin{figure}[t]
  \centering
  \includegraphics[width=0.98\columnwidth]
  {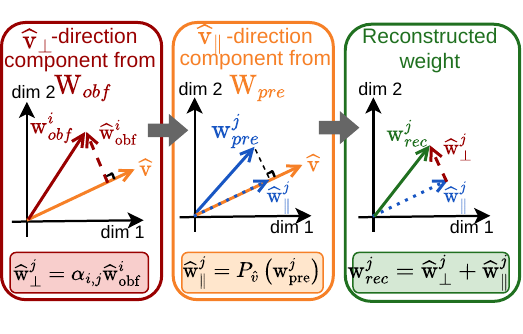}
  \caption{Two-component reconstruction in \attackname.  The projected
  exposed row provides the orthogonal component after scale correction,
  while the matched public row supplies only the missing component along
  $\widehat v$.}
  \label{fig:two_component_reconstruction}
\end{figure}

Projection removes the shared injection but also discards the victim's
legitimate component along $\widehat v$.  \attackname~therefore reconstructs
each matched victim row from two components, as illustrated in
Figure~\ref{fig:two_component_reconstruction}: an orthogonal component
recovered from the exposed row and a one-dimensional parallel component
completed from its matched public ancestor.

Consider an exposed row $y_i$ matched to public row $u_j$, where
$j=\widehat\pi(i)$. The attack treats $u_j$ as the public ancestor of the
unknown victim row $w_j$. After projection, $y_i$ approximately preserves
the direction of $w_j$ in the orthogonal subspace but remains affected by its
unknown scale. Following the public-weight-based scale correction used in
ArrowCloak~\cite{wang2025game}, we use the matched public row as the scaling
reference in the projected space. The corresponding rescaling factor is
\begin{equation}
\alpha_{ij}
=
\operatorname{sign}
\left[
\cos\!\left(
y_i\Pi_{\perp},
u_j\Pi_{\perp}
\right)
\right]
\frac{\|u_j\Pi_{\perp}\|_2}
{\|y_i\Pi_{\perp}\|_2}.
\label{eq:spectral_rescale}
\end{equation}
The resulting estimate of the victim row's orthogonal component is
\begin{equation}
\widehat w_{j,\perp}
=
\alpha_{ij}y_i\Pi_{\perp}.
\label{eq:spectral_orthogonal_component}
\end{equation}
Following the same public-reference principle, this correction uses the
projected public norm as an approximation to the unknown victim norm rather
than recovering the private scale $d_i$ exactly. Here, $\alpha_{ij}$ is an
approximate rescaling factor inferred from the matched public row; it is not
an estimate of $d_i^{-1}$.
The reconstructed component therefore retains the victim-specific direction
and residual information carried by the exposed row.

The scale correction above follows ArrowCloak's public-reference strategy,
but projection introduces an additional problem specific to \attackname:
the victim's legitimate component along $\widehat v$ is removed together
with the shared injection. This component cannot be separated from the
shared injection using a single exposed row. We therefore use only the
matched public row's one-dimensional projection along $\widehat v$ as a
completion term:
\begin{equation}
\widehat w_{j,\parallel}
=
u_j
\frac{\widehat v\widehat v^{\top}}
{\|\widehat v\|_2^2}.
\label{eq:spectral_parallel_component}
\end{equation}
The final reconstructed victim row combines the recovered and completed
components:
\begin{equation}
\widehat w_j
=
\widehat w_{j,\perp}
+
\widehat w_{j,\parallel}.
\label{eq:spectral_two_component}
\end{equation}

Thus, the public ancestor contributes only the one-dimensional information
lost during projection, while the remaining $d-1$ dimensional component is
recovered from the exposed victim row.  Section~\ref{sec:eval_spectralleak}
evaluates how closely these reconstructed weights match the victim and how
much model utility they recover.

\subsection{Surrogate Construction}
\label{sec:spectral_surrogate}

After reconstructing the protected rows, \attackname uses
$\widehat\pi$ to restore their model ordering and assembles the recovered
matrices in the victim architecture. Replacing the protected linear weights
with these matrices yields an initial reconstructed model $M_{\mathrm{rec}}$.
Small errors in correspondence, scale correction, and component completion may
accumulate across layers. We therefore initialize a surrogate
$M_{\mathrm{sur}}$ from $M_{\mathrm{rec}}$ and optionally fine-tune it using
at most $1\%$ of the public task training split. This refinement requires no
victim queries, victim outputs, or private training data.

Accordingly, \attackname targets \emph{functional recovery}: it reconstructs
weights from the exposed matrices and public ancestor, then uses limited
public data to recover residual utility. In contrast, Section~\ref{sec:blockhlp}
introduces a modular attack that reconstructs encoded victim weights without
queries, labels, or fine-tuning.

\subsection{Security Implications}
\attackname neither recovers secret obfuscation parameters nor compromises
the TEE. It uses only the transformed matrices intentionally exposed to the
accelerator. Reusing one hidden direction enables lightweight correction, but
also couples the exposed vectors through a shared rank-one relation. Thus,
ArrowCloak hides vector-level directional similarity, but the shared rank-one reuse remains observable at the matrix level.

\section{\blockhlpname: Breaking Modular ArrowCloak}
\label{sec:blockhlp}

Field-wide mod-$Q$ arithmetic suppresses the Euclidean PC1 signal exploited by \attackname, but it does not remove ArrowCloak's shared rank-one relation.
\blockhlpname targets this algebraic residue: it cancels the reused mask within small row blocks, recovers the victim row span of each block, and intersects overlapping spans to isolate individual encoded victim rows.

\subsection{Attack Target and Overview}
\label{sec:blockhlp_overview}
\blockhlpname is a new attack, whereas the finite-field target it analyzes follows the reduction to LWE problem formulation presented by the ArrowCloak~\cite{wang2025game}. We operationalize that formulation by applying fixed-point encoding and field-wide mod-\(Q\) arithmetic consistently to the exposed weights and all secret coefficients. This completion favors the defense and preserves the original diagonal-plus-rank-one structure.

Under this realization, the accelerator observes
\begin{equation}
Y=P(D+rk^{\top})W\pmod Q,
\label{eq:blockhlp_target}
\end{equation}
where $W$ is the fixed-point victim matrix, $D$ is diagonal, $P$ is a secret row permutation, and $rk^{\top}$ is the reused rank-one mask. The attacker knows $Y$ and the public system parameters but not $W$, $D$, $r$, $k$, or $P$.

The attack has three stages. First, it uses fixed-dimensional lattice reduction to recover the victim row span associated with each small exposed row block. Second, it arranges overlapping blocks so that pairwise span intersections isolate individual victim rows. Third, it validates every candidate using the complete accelerator-visible matrix. A public ancestor, when available, is used only after algebraic recovery to determine row signs and semantic positions.

\subsection{Recovering Hidden Row Spans}
\label{sec:blockhlp_spans}

Consider a block $I$ containing $b$ exposed rows. From
Section~\ref{sec:finite_field_leakage_view},
\begin{equation}
Y_I=D_IW_{\pi(I)}+r_Iv^{\top}\pmod Q,
\label{eq:blockhlp_block}
\end{equation}
where $W_{\pi(I)}$ contains the corresponding victim rows. Any blockwise coefficient vector
$c\in\mathbb F_Q^{b\times1}$ satisfying
$c^{\top}r_I=0$ cancels the shared term:
\begin{equation}
c^{\top}Y_I=\beta^{\top}W_{\pi(I)}\pmod Q, \beta=D_Ic.
\label{eq:block_combination}
\end{equation}
Thus, a short combination of exposed rows can become a short combination of victim rows even though neither the cancellation coefficients nor the victim rows are known.

The integer lifts of the unknown cancellation coefficients form a determinant-$Q$ lattice. \blockhlpname selects $t$ accelerator-visible coordinates from $Y_I$, constructs the corresponding Construction A lattice, and applies LLL~\cite{lenstra1982factoring}. Each returned combination is then applied to all $d$ coordinates and retained only if it passes the public consistency checks in Section~\ref{sec:blockhlp_validation}. Appendix~\ref{sec:appendix_blockhlp_details} gives the complete lattice construction and the short-vector separation conditions. Under those conditions, $b$ independent accepted combinations recover
\begin{equation}
S_I=\operatorname{span}_{\mathbb Q}
\left\{W_{\pi(i)}:i\in I\right\}.
\label{eq:blockhlp_row_span}
\end{equation}
Only the selected $t$ coordinates determine the lattice instance, but the accepted coefficients recover the span over all $d$ victim coordinates.

We use \(b=8\) and \(t=24\) throughout the evaluation. This fixed operating point balances span dimension, block coverage, and short-vector separation, while keeping every lattice instance 24-dimensional and independent of the full row width \(d\).
We therefore use $b=8,t=24$ unchanged across architectures rather than tuning them per model.

\subsection{Recovering Individual Victim Rows}
\label{sec:blockhlp_intersections}

The recovered space $S_I$ contains a block of victim rows but does not identify
them individually. Let exposed blocks $I$ and $J$ share exactly one index $i$.
If the victim rows associated with $I\cup J$ are linearly independent over
$\mathbb Q$, then
\begin{equation}
S_I\cap S_J=\operatorname{span}_{\mathbb Q}
\left(W_{\pi(i)}\right).
\label{eq:blockhlp_intersection}
\end{equation}
Indeed, a vector in the intersection has one expansion over $I$ and another over $J$. Subtracting the expansions creates a dependence over $I\cup J$; linear independence forces every coefficient outside $I\cap J=\{i\}$ to vanish. The attack checks that the intersection is one-dimensional and rejects the candidate otherwise.

Because $W$ uses a fixed-point integer encoding, exact rational intersection returns a projective integer row. Dividing by the coordinate gcd yields its primitive representative. If the encoded victim row is primitive, this representative equals the victim row up to sign. If its gcd is greater than one, the absolute integer scale remains ambiguous; we therefore restrict claims of exact encoded-row recovery to checkpoints that pass the primitive-row audit in Section~\ref{sec:eval_blockhlp}.

We realize the required overlaps with a grid schedule. For a block size $b$, place $b^2$ exposed rows in a $b\times b$ grid and recover the spans of its $b$ row blocks and $b$ column blocks. Each cell belongs to exactly one row block and one column block, so their intersection isolates the victim row at that cell. A group therefore needs $2b$ lattice reductions and $b^2$ span intersections. With our default $b=8$, each 64-row group uses 16 reductions and 64 intersections---this is why the implementation uses an $8\times8$ grid. More generally, a matrix with $m$ rows needs $2m/b$ reductions when $m$ is divisible by $b^2$; hence the evaluated 768- and 3,072-row matrices require 192 and 768 reductions, respectively. Every reduction remains 24-dimensional, independent of the full row width $d$.

\subsection{Validation and Row Labeling}
\label{sec:blockhlp_validation}

\blockhlpname validates recovery without a victim prediction oracle. First, each horizontal/vertical span pair must have a one-dimensional intersection. Second, coefficient relations learned from the selected coordinates must remain consistent over every accelerator-visible coordinate. If either check fails, the attacker discards the coordinate window and retries with a disjoint one. These checks and retries use no victim row, secret coefficient, correct permutation, task label, or victim output.

The algebraic stage returns an unordered set of signed primitive rows. When a public ancestor is available, a global one-to-one cosine assignment determines their signs and semantic positions. The public model contributes no new weight coordinate and does not determine integer magnitude; both come from algebraic recovery. Without a public ancestor, \blockhlpname still returns the unordered projective row set but cannot generally place those rows into an executable model. We therefore report algebraic row recovery and public-ancestor labeling separately.

\subsection{Scope and Guarantees}
\label{sec:blockhlp_scope}

\blockhlpname is not a general polynomial-time algorithm for arbitrary hidden lattice problems. It works well for the targeted models in our specific parameter settings. It relies on the codimension-one cancellation relation created by rank-one reuse, which decomposes full-model recovery into fixed-size lattice instances. Exact encoded-row recovery additionally requires separation between target and unrelated lattice vectors, independence of the overlapping row spans, and a known integer scale (primitive rows in our implementation).
We validate these conditions on the exact-recovery configurations reported in Section~\ref{sec:eval_blockhlp}.

The result therefore has a precise interpretation: field-wide arithmetic can hide the spectral manifestation of ArrowCloak's reused direction~\cite{wang2025game}, but the same reuse still creates publicly testable cancellation relations. Removing that relation, rather than merely changing its arithmetic representation, motivates the maximal-rank mitigation in the next section.

\section{\defensename: Removing Shared Rank-One Reuse}
\label{sec:butterfly_defense}

\attackname and \blockhlpname operate in different arithmetic domains, but they succeed for the same structural reason: ArrowCloak reuses one mask direction across every protected row. Its injected matrix $rv^{\top}$ has rank one. Over the reals, \attackname estimates the repeated direction and projects it away; over $\mod Q$, \blockhlpname finds nonzero row combinations that cancel the same repeated direction. Thus, changing the arithmetic hides one manifestation of the leakage without removing its common primitive.

A natural repair is to replace the rank-one mask with a rank-$\rho$ mask.
For constant $\rho$, a factored mask can still be corrected with
$O(\rho n\ell)$ trusted work for an $n\times n$ layer. However, this
only expands the leaked component into a $\rho$-dimensional principal
subspace. In our rank-2 and rank-3 ablations, projecting out the first two or three principal components restores the private–public row correspondence. Increasing the rank by a small constant therefore delays adaptive PCA rather than eliminating its primitive.
Resistance to such subspace removal requires the mask rank to scale to \(n\). A generic full-rank mask provides this protection but incurs \(O(n^{2}\ell)\) trusted correction, eliminating the computational benefit of offloading. \defensename resolves this tension through a keyed full-rank butterfly transform~\cite{dao2019butterfly}. It provides rank-\(n\) protection with \(O(n\ell\log n)\) trusted correction—only a logarithmic factor above ArrowCloak’s \(O(n\ell)\) cost and asymptotically below dense masking. For rectangular layers, it achieves the maximal rank \(\min(m,d)\) with \(O(\ell\max\{m,d\}\log d)\) trusted work.

\begin{table}[t]
\centering
\caption{Trusted correction for an $m\times d$ layer and $\ell$ input
activations. For a square layer with $m=d=n$, the three costs reduce to
$O(n\ell)$, $O(n\ell\log n)$, and $O(n^2\ell)$, respectively. Costs are
asymptotic and do not imply equal constants or measured TEE latency.}
\label{tab:defense_cost_ladder}
\begingroup
\footnotesize
\setlength{\tabcolsep}{3.3pt}
\resizebox{\columnwidth}{!}{%
\begin{tabular}{@{}l|c|c@{}}
\toprule
\textbf{Correction mask} & \textbf{Mask rank} & \textbf{TEE work} \\
\midrule
\sysname~ shared direction & $1$ & $O(\ell(m+d))$ \\
\defensename~ butterfly mask & $\min(m,d)$
& $O(\ell\max\{m,d\}\log d)$ \\
Independent dense mask & $\min(m,d)$ & $O(\ell md)$ \\
\bottomrule
\end{tabular}}
\endgroup
\end{table}

\subsection{Keyed Maximal-Rank Butterfly Mask}

For a row-oriented layer $W\in\mathbb R^{m\times d}$, \defensename samples an invertible diagonal matrix $D$, a row permutation $P$, a mask strength $\alpha$, and a keyed mask $M_{\theta}\in\mathbb R^{m\times d}$. It exposes
\begin{equation}
Y=P(DW+\alpha M_{\theta}).
\label{eq:butterfly_mask}
\end{equation}
The construction guarantees
\begin{equation}
\operatorname{rank}(M_{\theta})=\min(m,d).
\label{eq:butterfly_max_rank}
\end{equation}
Consequently, a square mask is full rank, while a rectangular mask has the largest rank allowed by its dimensions. Unlike ArrowCloak's $rv^{\top}$ term, the rows of $M_{\theta}$ are not scalar multiples of one reused direction.

For a square layer, $M_{\theta}=SZ_{\theta}$, where $S$ is a keyed sign diagonal and $Z_{\theta}$ is a product of $L=\lceil\log_2d\rceil$ sparse butterfly stages. Each stage pairs coordinates and applies independent invertible $2\times2$ rotations. Every stage is full rank, so their product is full rank. For $m<d$, the construction retains $m$ rows of a keyed $d\times d$ mask; for $m>d$, it stacks independently keyed $d\times d$ masks and truncates to $m$ rows. Appendix~\ref{sec:appendix_butterfly} gives the complete construction and rank argument.

\subsection{Trusted Correction and Cost}
\label{sec:butterfly_cost}

Suppose the accelerator receives $Y$ and returns
\begin{equation*}
O_{\mathrm{obf}}=XY^{\top},
\end{equation*}
where $X\in\mathbb R^{\ell\times d}$ contains $\ell$ input activations. From Equation~\eqref{eq:butterfly_mask}, the TEE obtains the desired layer output as
\begin{equation}
XW^{\top}=\left(O_{\mathrm{obf}}P-\alpha XM_{\theta}^{\top}\right)D^{-1}.
\label{eq:butterfly_recovery}
\end{equation}

Materializing $M_{\theta}$ does not make trusted correction efficient: a dense implementation of $XM_{\theta}^{\top}$ would still cost $O(\ell md)$. \defensename instead applies the keyed butterfly stages directly. Each stage contains disjoint $2\times2$ operations and costs linear work in its width. Accounting for rectangular layers, the trusted mask correction costs
\begin{equation}
O\!\left(\ell\max\{m,d\}\log d\right).
\label{eq:butterfly_complexity}
\end{equation}
This is a logarithmic factor above ArrowCloak's $O(\ell(m+d))$ correction and asymptotically below dense masking's $O(\ell md)$ cost. Table~\ref{tab:defense_cost_ladder} states both the general rectangular-layer costs and their square-layer shorthand.

\subsection{Why the Attacks Lose Their Primitive}
\label{sec:butterfly_attack_analysis}

\paragraph{\attackname.}
ArrowCloak exposes many rows containing the same direction $v$, creating a dominant rank-one component that PC1 can estimate. \defensename replaces that outer product with a maximal-rank mask whose rows span different directions.
There is therefore no single reused component whose projection restores every private/public row relation. Section~\ref{sec:eval_defense} verifies this claim against both PC1 removal and adaptive removal of multiple principal components.

\paragraph{ArrowMatch.}
Maximal rank alone does not guarantee that each exposed row is far from its public ancestor. The mask strength $\alpha$ is therefore chosen relative to the victim tensor's row norm so that each row receives a substantial, row-specific directional perturbation. We measure one-to-one correspondence recovery directly rather than treating rank as a proxy for resistance.

\paragraph{\blockhlpname.}
Within an ArrowCloak block, all mask rows are multiples of one vector, leaving a codimension-one family of nonzero coefficients that cancel the mask. For a \defensename block $I$ whose mask rows are linearly independent,
\begin{equation}
c^{\top}M_{\theta,I}=0\Longrightarrow c=0.
\label{eq:butterfly_no_cancellation}
\end{equation}
The nonzero cancellation space required by \blockhlpname is therefore absent. Appendix~\ref{sec:appendix_butterfly} gives the detailed argument, and Section~\ref{sec:eval_defense} tests the transferred attack on every protected BERT tensor.

\subsection{Security Scope}
\label{sec:butterfly_scope}

The analysis above explains why \defensename removes the concrete rank-one primitives used by \attackname and \blockhlpname. 

Increasing $\alpha$ strengthens directional hiding but also increases the trusted correction in Equation~\eqref{eq:butterfly_recovery}, creating a floating-point cancellation tradeoff. Our evaluation therefore measures correspondence recovery, adaptive spectral recovery, blockwise cancellation, output error, task accuracy, and trusted arithmetic cost. These experiments support \defensename as a mitigation against the attacks evaluated in this paper.

\section{Evaluation}
\label{sec:evaluation}

We answer four questions: (1) Does \attackname~ recover useful models from
the released real-valued scheme? (2) Which modular key distributions remove
the PC1? (3) Does \blockhlpname~ recover full models after
 field-wide mod-$Q$ protection? and (4) Can the root-cause repair in  Section~\ref{sec:butterfly_defense} suppress correspondence and spectral
 leakage, remove \blockhlpname's cancellation primitive, and preserve model
 utility?

\subsection{Experimental Setup}
\label{sec:experimental_setup}
We evaluate ViT-Base~\cite{dosovitskiy2021vit} on CIFAR-10,
CIFAR-100~\cite{krizhevsky2009cifar}, and Food-101~\cite{bossard2014food101};
BERT-Base~\cite{devlin2019bert} and GPT2-Base~\cite{radford2019gpt2} on MNLI,
QQP, SST-2, and QNLI from GLUE~\cite{wang2018glue}; and GPT2-XL on SST-2.
Together, these form 12 task settings across seven datasets. Victim models are
fine-tuned from public checkpoints, and all ArrowMatch values are our re-runs
against the same victims. We report end-to-end accuracy, correspondence
recovery, spectral energy, weight similarity, and exact encoded row recovery.
\attackname uses at most $1\%$ of the public task training split, whereas
\blockhlpname uses no victim queries, labels, or fine-tuning.

We additionally evaluate model scale using Qwen3-8B~\cite{yang2025qwen3},
Llama-3.1-8B~\cite{grattafiori2024llama3}, and
Gemma-4-E2B~\cite{gemmateam2026gemma4}, each fine-tuned on SST-2 as causal label
completion. We protect all two-dimensional attention and MLP projections,
covering 224--275 tensors, 0.94--1.40 million rows, and 1.86--6.98 billion
coordinates per model. These experiments use one fixed key seed and no
public-data refinement, so the reported \attackname accuracy reflects
reconstruction alone. Appendix~\ref{sec:appendix_experimental_setup} provides the complete checkpoints,
protected objects, and platform settings.

\subsection{RQ1: The Released Scheme}
\label{sec:eval_spectralleak}

\begin{table}[t]
\centering
\caption{End-to-end accuracy (\%) against released real-valued \sysname.
ArrowMatch is our re-run against the same victim checkpoints.}
\label{tab:attack_performance}
\begingroup
\footnotesize
\renewcommand{\arraystretch}{1.05}
\setlength{\tabcolsep}{2.2pt}
\begin{tabular}{@{}llrrrr@{}}
\toprule
\textbf{Model} & \textbf{Task} & \textbf{Exposed} & \makecell{\textbf{ArrowMatch}\\\textbf{re-run}}
& \textbf{\attackname} & \textbf{Victim} \\
\midrule
\multirow{3}{*}{ViT-Base}
 & C10  & 10.00 & 54.00 & \textbf{96.02} & 98.85 \\
 & C100 & 1.00  & 46.03 & \textbf{86.97} & 92.35 \\
 & F101 & 0.99  & 29.03 & \textbf{79.15} & 88.79 \\
\midrule
\multirow{4}{*}{BERT-Base}
 & MNLI & 32.74 & 45.88 & \textbf{83.42} & 83.69 \\
 & QQP  & 63.18 & 73.84 & \textbf{90.52} & 90.90 \\
 & SST2 & 50.92 & 73.85 & \textbf{90.83} & 91.86 \\
 & QNLI & 49.46 & 59.66 & \textbf{90.92} & 90.70 \\
\midrule
\multirow{4}{*}{GPT2-Base}
 & MNLI & 31.82 & 46.16 & \textbf{80.53} & 81.17 \\
 & QQP  & 37.07 & 74.06 & \textbf{86.43} & 87.23 \\
 & SST2 & 50.92 & 54.70 & \textbf{90.71} & 91.40 \\
 & QNLI & 50.50 & 54.35 & \textbf{85.67} & 86.62 \\
\midrule
GPT2-XL & SST2 & 49.08 & 59.06 & \textbf{94.61} & 94.61 \\
\midrule
\textbf{Average} & -- & 35.64 & 55.89 & \textbf{87.98} & 89.85 \\
\bottomrule
\end{tabular}
\endgroup
\end{table}

Table~\ref{tab:attack_performance} gives all per-task end-to-end
results. Directly running
the obfuscated models outside the TEE yields $35.64\%$ average accuracy, so
the public matrices are not trivially usable.  The prior direction-matching
attack re-run reaches $55.89\%$.  After removing the shared component,
\attackname~ reaches $87.98\%$, within 1.87 percentage points of the
$89.85\%$ white-box average.

The internal measurements explain the task result.  Projected assignment
recovers every correspondence for BERT, GPT2-Base, and GPT2-XL and at least
$99.89\%$ for ViT.  Reconstructed-weight cosine similarity ranges from
$98.15\%$ to $99.98\%$.  Appendix~\ref{sec:appendix_spectral} gives the
per-task values.  These measurements rule out an explanation based only on
fine-tuning: the correspondence and weights are already recovered before the
small-data refinement.

\subsection{RQ2: Spectral Leakage under Modular Protection}
\label{sec:eval_modq}

\begin{table}[t]
\centering
\caption{Arithmetic hardening on BERT/SST-2 ($Q=2^{31}-1$, $s=2^{16}$;
victim accuracy $91.6284\%$).}
\label{tab:modq_comparison}
\begingroup
\footnotesize
\setlength{\tabcolsep}{2.6pt}
\resizebox{\columnwidth}{!}{%
\begin{tabular}{@{}l|r|r|r|r@{}}
\toprule
\textbf{Instantiation} & \textbf{PC1 energy} & $|\cos(\mathrm{PC1},v)|$
& \textbf{Perm.} & \textbf{Pre-FT acc.} \\
\midrule
Real & 99.9924 & 100.0000 & 100.0000 & 91.7431 \\
mod-$Q$, small keys & 99.9923 & 100.0000 & 100.0000 & 91.6284 \\
mod-$Q$, full-field & 0.4048 & 2.5012 & 0.0808 & 50.9174 \\
\bottomrule
\end{tabular}}
\parbox{0.96\columnwidth}{\scriptsize Values are percentages. The maximum
wrap fraction is 0 for small keys and 1 for full-field keys.}
\endgroup
\end{table}

Table~\ref{tab:modq_comparison} isolates the effect of the key distribution.
Small keys produce no wrap and preserve the real spectral component.  With
field-wide keys, every tested tensor has wrapped coordinates, PC1 alignment
and permutation recovery become chance, and the unrecovered model stays at
the public-control accuracy.  Thus, the stronger arithmetic idea does stop
\attackname~ when field-wide randomness actually induces wrap; merely adding
a modulus to the released small keys does not.

The result generalizes.  Across ViT/CIFAR-10, CIFAR-100, Food-101,
GPT2-Base/SST-2, and GPT2-XL/SST-2, small keys retain at least $99.99\%$ PC1
energy and $99.89\%$ permutation recovery.  Full-field keys reduce PC1 energy
to $0.19$--$0.41\%$ and permutation recovery to $0.045$--$0.094\%$.
Appendix~\ref{sec:appendix_modq} reports each pair.

\begin{table*}[t]
\centering
\caption{Scale extension to modern language models fine-tuned on SST-2.
All values are percentages. ``Corr.'' denotes row correspondence recovery;
``proj. cos.'' denotes row-weighted projected cosine similarity before
refinement; ``post-PCA'' reports the best correspondence after removing
1, 8, or 32 exposed principal components.}
\label{tab:modern_llm_extension}
\begingroup
\scriptsize
\renewcommand{\arraystretch}{1.02}
\setlength{\tabcolsep}{5.2pt}
\resizebox{\textwidth}{!}{%
\begin{tabular}{@{}l|rrr@{}}
\toprule
\textbf{Metric}
& \textbf{Qwen3-8B}
& \textbf{Llama-3.1-8B}
& \textbf{Gemma-4-E2B} \\
\midrule

\multicolumn{4}{@{}l}{\textit{Released real-valued scheme}} \\
Victim accuracy
& 96.56 & 96.22 & 96.45 \\
Exposed accuracy
& 49.54 & 49.08 & 50.00 \\
ArrowMatch corr. / accuracy
& 9.096 / 50.34
& 9.080 / 90.48
& 9.151 / 49.77 \\
\attackname~PC1 energy / mask alignment
& 99.9984 / 100.0000
& 99.9984 / 100.0000
& 99.9969 / 100.0000 \\
\attackname~corr. / proj. cos.
& 99.606 / 99.885
& 100.000 / 99.932
& 99.996 / 99.951 \\
\attackname~accuracy (no refinement)
& 96.22 & 96.56 & 96.45 \\

\midrule
\multicolumn{4}{@{}l}{%
\textit{\defensename~at $256\times$ the median victim-row norm}} \\
PC1 energy / raw corr. / post-PCA corr.
& 0.0454 / 0.2051 / 0.2033
& 0.0464 / 0.0432 / 0.0424
& 0.1835 / 0.0550 / 0.0536 \\
Protected / victim accuracy
& 96.67 / 96.56
& 96.45 / 96.22
& 96.22 / 96.45 \\
Butterfly / dense operations (\%)
& 0.516 & 0.544 & 1.278 \\

\bottomrule
\end{tabular}%
}
\endgroup
\end{table*}

\subsection{RQ3: Recovery under Mod-$Q$ Protection}
\label{sec:eval_blockhlp}

On the BERT anchor, \blockhlpname exactly recovers all 72 protected tensors,
comprising 82,944 rows and 84,934,656 fixed-point coordinates, in 233.44
seconds including evaluation. The reconstructed model retains the victim's
$91.6284\%$ task accuracy. Appendix Table~\ref{tab:blockhlp_models} gives the
cross-architecture audit: GPT2-Base is recovered exactly on both tasks, while
the few imperfect rows in ViT and GPT2-XL leave task accuracy effectively
unchanged. Appendix~\ref{sec:appendix_ablations} provides the per-shape
recovery diagnostics, including short-vector scales, candidate gaps, and
coordinate-window retries.

\subsection{RQ4: Attack Mitigation with \defensename}
\label{sec:eval_defense}
With independent tensor keys and a mask strength of $512\times$ the median
victim-row norm, full-model BERT protection preserves the victim's
$91.6284\%$ accuracy while reducing PC1 energy and raw correspondence
recovery to $0.13049\%$ and $0.11695\%$, respectively. The transferred
\blockhlpname implementation recovers none of the 82,944 victim rows.
Appendix Tables~\ref{tab:butterfly_defense}
and~\ref{tab:butterfly_blockhlp} provide the complete numerical diagnostics
and attack-transfer results, respectively.

\subsection{Scale Extension to Modern LLMs}
\label{sec:modern_llm_extension}

We evaluate whether the observed leakage and the proposed mitigation extend
to Qwen3-8B, Llama-3.1-8B, and Gemma-4-E2B. Unlike the preceding experiments,
this evaluation uses no public-data refinement; the reported \attackname
accuracies therefore measure reconstruction alone.

As shown in Table~\ref{tab:modern_llm_extension}, the shared-direction
leakage persists at this scale. \attackname recovers
$99.606\%$--$100\%$ of row correspondences and reconstructs models within
$0.34$ percentage points of victim accuracy. In contrast, ArrowMatch recovers
only $9.080\%$--$9.151\%$ of correspondences. Its $90.48\%$ accuracy on
Llama-3.1-8B should not be interpreted as successful parameter recovery:
correspondence recovery remains $9.080\%$, and the functional result largely
comes from the public ancestor.

\defensename reduces PC1 energy to $0.0454\%$--$0.1835\%$ and keeps both
direct and adaptive PCA correspondence recovery below $0.206\%$. Protected
accuracy differs from victim accuracy by at most two of the 872 validation
examples. Butterfly correction requires only $0.516\%$--$1.278\%$ of the
operations of the protected dense layers. These values are arithmetic
operation counts rather than measured TEE latency. Appendix Tables~\ref{tab:modern_model_audit}
and~\ref{tab:modern_butterfly_diagnostics} provide the protected-object audit and detailed numerical results.

\section{Conclusion}

This paper studies the security of matrix-level weight obfuscation in TEE-shielded model partitioning and identifies shared rank-one reuse as a fundamental source of information leakage. We propose \attackname and \blockhlpname to exploit this structure over the real numbers and a finite field, respectively. Our experiments further show that the real-valued leakage persists in Qwen3-8B, Llama-3.1-8B, and Gemma-4-E2B, covering models with 5.1--8.2 billion parameters. To mitigate this leakage, we introduce \defensename, a keyed maximal-rank butterfly mask that eliminates shared low-rank reuse while preserving the measured model utility. Our evaluation demonstrates that \defensename suppresses the evaluated spectral and correspondence attacks at this scale, with an $O(\ell\max\{m,d\}\log d)$ trusted-side correction for an $m\times d$ linear layer. More broadly, our findings show that lightweight trusted correction should avoid reusing hidden directions across an exposed matrix. Future defenses should consider both per-vector concealment and matrix-level leakage when protecting accelerator-visible weights.

\bibliographystyle{plain}
\bibliography{ref}

\appendix
\section{Ethical Considerations}
This work evaluates the confidentiality of a research prototype for
TEE-assisted model deployment.  All experiments use models we trained from
public checkpoints and public datasets; they contain no personal or user
data.  The attacks begin only after an adversary has obtained the obfuscated
weights that the design intentionally releases to an untrusted accelerator,
and they do not bypass a TEE or another hardware isolation mechanism.  The
main dual-use risk is that the analysis could assist extraction of a model
protected by a similar scheme.  We mitigate that risk by reporting the design
root cause and defense tradeoffs, avoiding attacks on third-party deployments,
and preparing coordinated disclosure to the authors before public artifact
release.  To our knowledge, ArrowCloak is not deployed in a production system.

\section{Detailed Spectral Diagnostics}
\label{sec:appendix_spectral}
\paragraph{Principal component analysis.}
Given a matrix $X\in\mathbb R^{n\times d}$ whose rows are the input
vectors, let $\mu\in\mathbb R^{d\times1}$ denote the mean vector in
column form. Its centered form is
\begin{equation*}
    X_c=X-\mathbf{1}\mu^{\top},
\end{equation*}
where $\mathbf{1}\in\mathbb R^{n\times1}$.
The first principal component is
\begin{equation*}
u_1=
\operatorname*{arg\,max}_{\|u\|_2=1}
\|X_{\mathrm c}u\|_2^2,
\end{equation*}
or equivalently the leading right singular vector of $X_{\mathrm c}$.
Unlike a per-vector analysis, PCA identifies the dominant variation shared
across the complete collection of rows. We use this standard estimator to
test whether shared-direction injection produces a prominent component in
the accelerator visible matrix.

\begin{figure}[t]
  \centering
  \includegraphics[width=0.88\columnwidth]{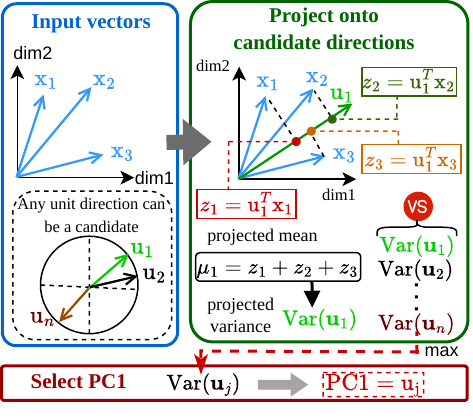}
  \caption{Visual intuition for PC1: among candidate directions, it maximizes
  the variance of the centered row projections.  This standard estimator is
  used by \attackname; the vulnerability is the protocol's repeated component,
  not PCA itself.}
  \label{fig:pca_pc1_toy}
\end{figure}

Tables~\ref{tab:spectral_diagnostics} and~\ref{tab:spectral_recovery}
consolidate the internal measurements
behind Section~\ref{sec:eval_spectralleak}.  The extracted PC1 aligns with the
true injected direction at $100.00\%$ to reported precision in every row.
Removing a random direction leaves the cosine distance equal to
$D_{\mathrm{ori}}$ to four decimals; it is omitted for space.
Figure~\ref{fig:pca_pc1_toy} gives the geometric intuition for the PC1
estimator used in these measurements.

\begin{table}[t]
\centering
\caption{Per-task spectral diagnostics.  $\rho_1$ is PC1 energy and
$d_{\mathrm{cos}}$ is cosine distance.}
\label{tab:spectral_diagnostics}
\begingroup
\footnotesize
\setlength{\tabcolsep}{2.5pt}
\resizebox{\columnwidth}{!}{%
\begin{tabular}{@{}l|l|r|r|r|r@{}}
\toprule
\textbf{Model} & \textbf{Task} & $\rho_1(W)$ & $\rho_1(Y)$
& $d_{\mathrm{ori}}$ & $d_{\mathrm{PC1}}$ \\
\midrule
\multirow{3}{*}{ViT-Base}
 & C10 & 2.27 & 99.99 & 0.8992 & 0.0011 \\
 & C100 & 2.27 & 99.99 & 0.8992 & 0.0013 \\
 & Food-101 & 2.27 & 99.99 & 0.8992 & 0.0016 \\
\midrule
\multirow{4}{*}{BERT-Base}
 & MNLI & 1.87 & 99.99 & 0.8993 & 0.0022 \\
 & QQP & 1.87 & 99.99 & 0.8994 & 0.0033 \\
 & SST2 & 1.87 & 99.99 & 0.8992 & 0.0011 \\
 & QNLI & 1.87 & 99.99 & 0.8992 & 0.0016 \\
\midrule
\multirow{4}{*}{GPT2-Base}
 & MNLI & 2.80 & 99.99 & 0.8998 & 0.0017 \\
 & QQP & 2.80 & 99.99 & 0.8997 & 0.0015 \\
 & SST2 & 2.80 & 99.99 & 0.8997 & 0.0014 \\
 & QNLI & 2.80 & 99.99 & 0.8997 & 0.0015 \\
\midrule
GPT2-XL & SST2 & 1.44 & 100.00 & 0.9042 & 0.0005 \\
\bottomrule
\end{tabular}}
\endgroup
\end{table}

\begin{table}[t]
\centering
\caption{Per-task correspondence recovery and reconstructed-weight cosine
similarity (\%).}
\label{tab:spectral_recovery}
\begingroup
\footnotesize
\setlength{\tabcolsep}{4.0pt}
\begin{tabular}{@{}l|l|r|r@{}}
\toprule
\textbf{Model} & \textbf{Task} & \textbf{Match} & \textbf{Similarity} \\
\midrule
\multirow{3}{*}{ViT-Base}
 & C10 & 99.90 & 98.96 \\
 & C100 & 99.89 & 98.77 \\
 & Food-101 & 99.90 & 98.15 \\
\midrule
\multirow{4}{*}{BERT-Base}
 & MNLI & 100.00 & 99.81 \\
 & QQP & 100.00 & 99.72 \\
 & SST2 & 100.00 & 99.93 \\
 & QNLI & 100.00 & 99.86 \\
\midrule
\multirow{4}{*}{GPT2-Base}
 & MNLI & 100.00 & 99.93 \\
 & QQP & 100.00 & 99.96 \\
 & SST2 & 100.00 & 99.98 \\
 & QNLI & 100.00 & 99.97 \\
\midrule
GPT2-XL & SST2 & 100.00 & 99.97 \\
\bottomrule
\end{tabular}
\endgroup
\end{table}

\section{Finite-Field Extension Results}
\label{sec:appendix_modq}

Table~\ref{tab:modq_extension} repeats the small-key/full-field comparison on
the extension checkpoints.  Small keys preserve the spectral leak on every
architecture; full-field keys suppress it on every architecture.

\begin{table}[t]
\centering
\caption{Small-key / full-field results for the modular extension models.
All entries are percentages.}
\label{tab:modq_extension}
\begingroup
\footnotesize
\setlength{\tabcolsep}{2.5pt}
\resizebox{\columnwidth}{!}{%
\begin{tabular}{@{}l|l|r|r|r@{}}
\toprule
\textbf{Model} & \textbf{Task} & \textbf{PC1 energy}
& \textbf{Perm. rec.} & \textbf{Pre-FT acc.} \\
\midrule
ViT-Base & C10 & 99.9925 / 0.4038 & 99.8951 / 0.0784 & 98.61 / 10.00 \\
ViT-Base & C100 & 99.9925 / 0.4053 & 99.8915 / 0.0940 & 91.33 / 0.74 \\
ViT-Base & Food-101 & 99.9925 / 0.4052 & 99.8855 / 0.0832 & 85.89 / 1.00 \\
GPT2-Base & SST2 & 99.9932 / 0.4063 & 100.0000 / 0.0796 & 91.8578 / 52.0642 \\
GPT2-XL & SST2 & 99.9984 / 0.1940 & 100.0000 / 0.0447 & 93.2339 / 49.0826 \\
\bottomrule
\end{tabular}}
\endgroup
\end{table}

\section{\blockhlpname Exact-Recovery Diagnostics}
\label{sec:appendix_ablations}

\paragraph{Per-shape behavior.}
Table~\ref{tab:blockhlp_shapes} separates the BERT anchor by tensor shape.
All public validations pass; one tall tensor needs a second coordinate window.

\begin{table}[t]
\centering
\caption{Per-shape BERT results (eight workers).}
\label{tab:blockhlp_shapes}
\begingroup
\footnotesize
\setlength{\tabcolsep}{2.8pt}
\resizebox{\columnwidth}{!}{%
\begin{tabular}{@{}l|r|r|r|r|r@{}}
\toprule
\textbf{Shape} & \textbf{Tensors} & \textbf{Exact rows}
& \textbf{Avg. s} & \textbf{Min gap} & \textbf{Retries} \\
\midrule
$768\times768$ & 48 & 36,864 / 36,864 & 12.27 & 3,864$\times$ & 0 \\
$768\times3072$ & 12 & 9,216 / 9,216 & 28.73 & 4,876$\times$ & 0 \\
$3072\times768$ & 12 & 36,864 / 36,864 & 48.14 & 3,579$\times$ & 1 \\
\midrule
\textbf{Total} & 72 & 82,944 / 82,944 & -- & -- & 1 \\
\bottomrule
\end{tabular}}
\endgroup
\end{table}

\section{More Details on the \blockhlpname Reduction}
\label{sec:appendix_blockhlp_details}

\paragraph{The hidden coefficient lattice.}
For a block $I$ of $b$ exposed rows, define
\begin{equation*}
a_I=D_I^{-1}r_I.
\end{equation*}
The integer coefficient vectors that cancel the shared component belong to
\begin{equation*}
L_I
=
\left\{
\beta\in\mathbb Z^b:
\beta^{\top}a_I=0\pmod Q
\right\}.
\end{equation*}
For nonzero $a_I$, consider the homomorphism
\begin{equation*}
\phi_I:\mathbb Z^b\rightarrow\mathbb F_Q,
\qquad
\phi_I(\beta)=\beta^{\top}a_I\bmod Q.
\end{equation*}
The map is surjective and $L_I=\ker\phi_I$. Consequently, $L_I$ is a
full-rank integer lattice in $\mathbb Z^b$ with index and determinant $Q$.
Although the cancellation space over $\mathbb F_Q$ has dimension $b-1$,
its integer preimage has rank $b$.

\paragraph{Expected short-vector scale.}
The Gaussian heuristic gives a characteristic short-vector scale of
$Q^{1/b}$. With $b=8$ and $Q=2^{31}-1$, this scale is approximately
$14.7$. Minkowski's theorem guarantees a nonzero vector with norm bounded
on the order of $\sqrt{b}Q^{1/b}$, but neither statement guarantees that
$b$ independent cancellation vectors appear in an LLL-reduced basis.
The attack therefore treats LLL outputs as candidates and validates them
using only public information.

\paragraph{Public Construction A lattice.}
The attacker cannot construct $L_I$ directly because $a_I$ is secret.
Instead, it chooses a public coordinate set $T\subseteq[d]$ of size $t$ and
forms
\begin{equation*}
\mathcal C_{I,T}
=
\operatorname{rowspan}_{\mathbb F_Q}(Y_{I,T})
\subseteq\mathbb F_Q^t,
\end{equation*}
where $Y_{I,T}$ restricts $Y_I$ to the selected coordinates. The associated
Construction A lattice is
\begin{equation*}
\Lambda_{I,T}
=
\left\{
z\in\mathbb Z^t:
z\bmod Q\in\mathcal C_{I,T}
\right\}.
\end{equation*}
Every modular combination of the selected exposed rows has an integer lift
in $\Lambda_{I,T}$. If a coefficient vector cancels the shared component,
then
\begin{equation*}
c^{\top}Y_{I,T}
=
\beta^{\top}W_{\pi(I),T}
\pmod Q.
\end{equation*}
Because the fixed-point victim weights and relevant integer coefficients are
small relative to $Q$, their centered lifts can be substantially shorter
than ordinary modular codewords.

\paragraph{Conditional span recovery.}
Suppose lattice reduction returns $b$ independent lifted codewords
$z_1,\ldots,z_b$ associated with cancellation coefficients
$\beta_1,\ldots,\beta_b\in L_I$, without centered wrap. Let
\begin{equation*}
B=
\begin{bmatrix}
\beta_1^{\top}\\
\vdots\\
\beta_b^{\top}
\end{bmatrix}.
\end{equation*}
If $B$ is nonsingular over $\mathbb Q$, applying the recovered public
row-code coefficients to all coordinates gives
\begin{equation*}
Z=BW_{\pi(I)}.
\end{equation*}
It follows that
\begin{equation*}
\operatorname{rowspan}_{\mathbb Q}(Z)
=
\operatorname{rowspan}_{\mathbb Q}(W_{\pi(I)})
=
S_I.
\end{equation*}
This implication is exact. The empirical step is whether LLL separates the
required cancellation combinations from unrelated lattice vectors.

\paragraph{Public acceptance checks.}
The implementation requires the selected candidates to be linearly
independent, measures their separation from the first rejected LLL
candidate, applies the recovered coefficients to complete exposed rows, and
checks that overlapping spans have the expected one-dimensional
intersections. If a check fails, it retries with a disjoint public coordinate
window. These checks require no victim weights, secret parameters, correct
permutation, labels, or model predictions.

They do not constitute a general certificate that every accepted projective
row equals a victim row. 
Exact equality is measured separately by the
challenger-only element-wise audit reported in
Section~\ref{sec:eval_blockhlp}.

\section{ButterflyCloak Construction and Analysis}
\label{sec:appendix_butterfly}

\noindent\textbf{Square masks.}
For a square layer with $m=d$, \defensename constructs
\begin{equation*}
M_{\theta}=SZ_{\theta},
\qquad
Z_{\theta}=B_LB_{L-1}\cdots B_1,
\qquad
L=\lceil\log_2 d\rceil,
\end{equation*}
where $S$ is a keyed random sign diagonal matrix. Each stage $B_j$ pairs
coordinates according to the key and applies independent invertible
$2\times2$ rotations. Every stage is full rank, so their product
$Z_{\theta}$ is full rank. Multiplication by $S$ preserves rank, giving
\begin{equation*}
\operatorname{rank}(M_{\theta})=d.
\end{equation*}

The TEE does not store or apply $M_{\theta}$ as a dense matrix. It stores the
stage pairings, signs, and rotation parameters and applies the stages
sequentially.

\noindent\textbf{Rectangular masks.}
When $m\leq d$, the construction retains the first $m$ rows of a keyed
$d\times d$ butterfly mask. When $m>d$, it stacks
$\lceil m/d\rceil$ independently keyed $d\times d$ masks and truncates the
result to $m$ rows. Under the stated independent-key construction, the
resulting mask satisfies
\begin{equation*}
\operatorname{rank}(M_{\theta})=\min(m,d).
\end{equation*}
This covers the $768\times3072$ and $3072\times768$ matrices evaluated in
our experiments.

\noindent\textbf{Effect on \attackname.}
ArrowCloak injects an outer-product term $rv^{\top}$, causing many exposed
rows to share one dominant direction. \attackname estimates this component
with PC1 and projects it away. In \defensename, the injected term is instead
$\alpha M_{\theta}$, whose rows span a maximal-rank space. There is therefore
no single shared direction whose removal restores all exposed rows to their
public ancestor directions.

\noindent\textbf{Effect on ArrowMatch.}
Removing rank-one reuse alone does not prevent vector matching if each
exposed row remains close to its public ancestor. \defensename therefore
chooses $\alpha$ so that each exposed row contains a substantial contribution
from a distinct mask row. The evaluation measures correspondence recovery
directly rather than assuming that maximal rank alone defeats ArrowMatch.

\noindent\textbf{Effect on \blockhlpname.}
For an ArrowCloak block, all mask rows are scalar multiples of one vector,
creating a codimension-one family of coefficients that cancel the mask.
For a \defensename block $I$ whose mask rows are linearly independent,
\begin{equation*}
c^{\top}M_{\theta,I}=0
\end{equation*}
implies $c=0$. The nonzero cancellation space required by \blockhlpname is
therefore absent.

\noindent\textbf{Security scope.}
These arguments explain why \defensename removes the primitives exploited by
the attacks evaluated in this work; they do not provide a cryptographic
confidentiality proof. The butterfly family remains structured, and we do
not rule out attacks that jointly estimate public weights, diagonal scales,
the hidden permutation, and butterfly parameters. Increasing $\alpha$ also
strengthens directional hiding while increasing floating-point cancellation
during trusted correction. We therefore evaluate \defensename as a systems
mitigation rather than a cryptographic endpoint.

\noindent\textbf{Complete cross-architecture recovery audit.}
Table~\ref{tab:blockhlp_models} gives the complete \blockhlpname~audit
referenced in Section~\ref{sec:eval_blockhlp}.

\begin{table*}[t]
\centering
\caption{\blockhlpname~ against field-wide mod-$Q$ protection. ``Exact rows'' is a
challenger-only element-wise audit after public-reference sign and position
assignment. Public checks accept every reported tensor but do not themselves
guarantee an exact encoded row.}
\label{tab:blockhlp_models}
\begingroup
\footnotesize
\renewcommand{\arraystretch}{1.08}
\setlength{\tabcolsep}{5.2pt}
\begin{tabular}{@{}l|l|r|r|r|r@{}}
\toprule
\textbf{Model} & \textbf{Task} & \textbf{Protected tensors}
& \textbf{Exact rows} & \textbf{Max integer error}
& \textbf{Recovered / victim acc. (\%)} \\
\midrule
ViT-Base & C10 & 72 & 82,876 / 82,944 & 7,905 & 98.70 / 98.69 \\
ViT-Base & C100 & 72 & 82,884 / 82,944 & 7,888 & 91.66 / 91.68 \\
ViT-Base & Food-101 & 72 & 82,884 / 82,944 & 7,787 & 88.51 / 88.51 \\
BERT-Base & SST2 & 72 & 82,944 / 82,944 & 0 & 91.6284 / 91.6284 \\
GPT2-Base & SST2 & 72 & 82,944 / 82,944 & 0 & 91.06 / 91.06 \\
GPT2-Base & QNLI & 72 & 82,944 / 82,944 & 0 & 87.75 / 87.75 \\
GPT2-XL & SST2 & 288 & 691,192 / 691,200 & 10,022 & 92.7752 / 92.7752 \\
\bottomrule
\end{tabular}
\endgroup
\end{table*}

Tables~\ref{tab:butterfly_defense} and~\ref{tab:butterfly_blockhlp} give the
full-BERT mitigation and transfer details referenced in
Section~\ref{sec:eval_defense}.

\begin{table}[t]
\centering
\caption{Full BERT/SST-2 \defensename~ evaluation over 72 tensors and 82,944
exposed rows. Entries except accuracy are percentages.}
\label{tab:butterfly_defense}
\begingroup
\footnotesize
\setlength{\tabcolsep}{2.1pt}
\resizebox{\columnwidth}{!}{%
\begin{tabular}{@{}r|r|r|r|r|r|r@{}}
\toprule
$\alpha$ & \textbf{PC1} & \textbf{Raw match} & \textbf{Rm. 1}
& \textbf{Rm. 8} & \textbf{Rm. 32} & \textbf{Accuracy} \\
\midrule
$256\times$ & 0.13065 & 0.13744 & 0.13383 & 0.10007 & 0.10971 & 91.6284 \\
$512\times$ & 0.13049 & 0.11695 & 0.11454 & 0.07595 & 0.09283 & 91.6284 \\
\bottomrule
\end{tabular}}
\parbox{0.96\columnwidth}{\scriptsize ``Rm.'' projects both exposed and public
rows away from the indicated number of exposed principal components. The
victim accuracy is $91.6284\%$; maximum sampled float32 output error is
$0.00338$.}
\endgroup
\end{table}

\begin{table}[t]
\centering
\caption{Full-BERT \blockhlpname~ transfer stress test against
$512\times$ \defensename. ``Public checks'' are internal consistency tests,
not an exact-recovery oracle.}
\label{tab:butterfly_blockhlp}
\begingroup
\footnotesize
\setlength{\tabcolsep}{4.0pt}
\begin{tabular}{@{}l|r@{}}
\toprule
\textbf{Metric} & \textbf{Result} \\
\midrule
Real mask blocks with row rank 8 & 20,736 / 20,736 \\
Tensors passing public checks & 28 / 72 \\
Exact victim rows (up to sign) & 0 / 82,944 \\
Primitive exposed rows returned & 6,876 / 82,944 \\
Wall-clock time (8 workers) & 183.54 s \\
\bottomrule
\end{tabular}
\endgroup
\end{table}

The transfer's 28 internally consistent outputs contain no exact victim row;
44 tensors reject all 16 public coordinate windows. For 6,876 rows the output
is instead the primitive form of the already public, mask-dominated exposed
row. The compact correction performs 3,538,944 rotation scalar operations per
token versus 169,869,312 for the protected dense layers ($2.083\%$). This count
excludes permutation, scaling, and memory traffic.

\section{Experimental Setup}
\label{sec:appendix_experimental_setup}

\noindent\textbf{Models and datasets.}
Following the original evaluation, we use ViT-Base on CIFAR-10, CIFAR-100,
and Food-101; BERT-Base and GPT2-Base on MNLI, QQP, SST-2, and QNLI; and
GPT2-XL on SST-2. All victim models are fine-tuned from public checkpoints.
The real-valued and modular experiments use their corresponding trained
checkpoints, and each result table includes the paired victim control.
All ArrowMatch results are our reruns on these checkpoints.

\noindent\textbf{Attack parameters.}
Our modular anchor is BERT-Base on SST-2 with seed 42 and 872 validation
examples. We use a fixed-point exponent of $f=16$, corresponding to the scale
$s=2^{16}$, and modulus $Q=2^{31}-1$. \attackname fine-tunes its
reconstructed surrogate for three epochs using 673 public examples, less than
$1\%$ of the training split, with learning rate $10^{-5}$ and batch size 32.
\blockhlpname uses no task labels or fine-tuning.

\noindent\textbf{Metrics and audit.}
We report task accuracy, correspondence recovery, PC1 energy, and weight
cosine similarity. For \blockhlpname, exact recovery means element-wise
equality with the defender's fixed-point integer matrix. The recovery code
receives only one accelerator-visible residue matrix and the public system
parameters. Private weights, keys, and permutations are loaded only by a
separate challenger after recovery for evaluation. Cached results are bound
to the checkpoint and complete attack configuration; reported BERT and
GPT2-XL runtimes are uncached.

\noindent\textbf{Platforms.}
Real-valued training and evaluation use two NVIDIA A6000 GPUs. Modular
experiments use RTX 4090 Laptop and RTX 4080 GPUs for model evaluation and
an NVIDIA H20-3e GPU for GPT2-XL training. \blockhlpname itself is CPU-only
and uses eight workers with exact SymPy arithmetic for lattice reduction and
span intersection.

\subsection{Modern LLM Extension}
\label{sec:appendix_modern_llm}

\noindent\textbf{Checkpoints and training.}
The public checkpoints are \texttt{Qwen/Qwen3-8B-Base} revision
\texttt{49e3418fbbbca6ecbdf9\allowbreak608b4d22e5a407081db4},
\texttt{meta-llama/Llama-3.1-8B} revision
\texttt{d04e592bb4f6aa9cfee9\allowbreak1e2e20afa771667e1d4b}, and
\texttt{google/gemma-4-E2B} revision
\texttt{d29ff6b45f081a49ee27\allowbreak33a859c9c9c2d95d1a6f}.
All checkpoints were verified against their reported revision hashes. The
three training runs complete at optimizer step 25,257, and the resulting
checkpoints are used throughout the corresponding attack and defense
experiments.

All models use the prompt \texttt{Review: <sentence> Sentiment:} and predict
a single \texttt{negative} or \texttt{positive} label token. We fine-tune for
three epochs with seed 42, learning rate $10^{-5}$, weight decay 0.01, warmup
ratio 0.06, sequence length 128, batch size 8, bfloat16, and gradient
checkpointing. Qwen and Llama update all parameters using 8-bit AdamW. Gemma
updates its text backbone and language model head using fused AdamW, while
its unused vision and audio towers remain frozen.
Table~\ref{tab:modern_model_audit} reports the exact parameter and protected
object counts.

\begin{table*}[t]
\centering
\caption{Checkpoint and protected object audit for the modern language
models. Parameter, row, and coordinate counts are exact; accuracy is measured
on all 872 SST-2 validation examples.}
\label{tab:modern_model_audit}
\begingroup
\footnotesize
\setlength{\tabcolsep}{5.0pt}
\begin{tabular}{@{}l|r|r|r|r|r|r@{}}
\toprule
\textbf{Model} &
\textbf{Total params.} &
\textbf{Trainable params.} &
\textbf{Tensors} &
\textbf{Rows} &
\textbf{Coordinates} &
\textbf{Victim acc. (\%)} \\
\midrule
Qwen3-8B &
8,190,735,360 &
8,190,735,360 &
252 &
1,400,832 &
6,945,767,424 &
96.5596 \\
Llama-3.1-8B &
8,030,261,248 &
8,030,261,248 &
224 &
1,376,256 &
6,979,321,856 &
96.2156 \\
Gemma-4-E2B &
5,104,297,504 &
4,628,569,344 &
275 &
941,312 &
1,863,057,408 &
96.4450 \\
\bottomrule
\end{tabular}
\endgroup
\end{table*}

\begin{table*}[t]
\centering
\caption{\defensename diagnostics on modern language models at a mask
strength of $256\times$ the median victim row norm. ``Rm. $k$'' denotes
correspondence recovery after removing the first $k$ exposed principal
components. Errors report the median and 95th percentile of tensor mean
absolute output error, followed by the maximum sampled error.}
\label{tab:modern_butterfly_diagnostics}
\begingroup
\scriptsize
\setlength{\tabcolsep}{3.5pt}
\begin{tabular}{@{}l|r|rrrr|r|r|r@{}}
\toprule
\textbf{Model} &
\textbf{PC1 energy (\%)} &
\multicolumn{4}{c|}{\textbf{Correspondence recovery (\%)}} &
\textbf{Mean error med. / p95} &
\textbf{Max error} &
\textbf{Protected / victim acc. (\%)} \\
\cmidrule(lr){3-6}
& &
\textbf{Raw} &
\textbf{Rm. 1} &
\textbf{Rm. 8} &
\textbf{Rm. 32} &
& & \\
\midrule
Qwen3-8B &
0.04542 &
0.20509 &
0.20331 &
0.19453 &
0.18289 &
$1.43{\times}10^{-4}$ / $4.49{\times}10^{-4}$ &
$6.03{\times}10^{-3}$ &
96.6743 / 96.5596 \\
Llama-3.1-8B &
0.04638 &
0.04316 &
0.04243 &
0.04149 &
0.04076 &
$7.65{\times}10^{-5}$ / $2.27{\times}10^{-4}$ &
$2.58{\times}10^{-3}$ &
96.4450 / 96.2156 \\
Gemma-4-E2B &
0.18349 &
0.05503 &
0.05365 &
0.04642 &
0.05184 &
$5.28{\times}10^{-5}$ / $1.47{\times}10^{-4}$ &
$1.53{\times}10^{-3}$ &
96.2156 / 96.4450 \\
\bottomrule
\end{tabular}
\endgroup
\end{table*}

\noindent\textbf{Protected tensors.}
Rows of each protected matrix are treated as weight vectors. For Qwen and
Llama, we protect every two-dimensional
\texttt{q/k/v/o/gate/up/down\_proj} matrix in the text transformer. For Gemma,
we protect the corresponding attention and MLP projections together with
\texttt{per\_layer\_input\_gate} and
\texttt{per\_layer\_projection}. Table~\ref{tab:modern_model_audit} reports
the resulting tensor, row, and coordinate counts, while the complete shape
histograms are retained in the machine-readable experiment manifests.
Embeddings, normalization parameters, and output heads remain unchanged,
consistent with the protected layer scope used in the GPT2 experiments.

\noindent\textbf{Keys, seeds, and attack settings.}
All ArrowCloak and \defensename runs use key seed 42, with an independently
derived random generator for each tensor. The ArrowCloak experiments use its
original real-valued coefficient distributions. \defensename uses a mask
strength of $256\times$ the median victim row norm, and the adaptive PCA test
removes 1, 8, or 32 exposed principal components.

Cached results are bound to the algorithm version, checkpoint, tensor,
arithmetic parameters, seed, and attack configuration. The machine-readable
manifests record the corresponding SHA-256 checksums.

\noindent\textbf{Software and hardware.}
Training and evaluation use PyTorch 2.13.0+cu130, Transformers 5.15.1,
Datasets 5.0.1, and bitsandbytes 0.50.1 on a single NVIDIA H20-3e GPU. Peak
allocated training memory is 53.198, 51.645, and 39.659 GiB for Qwen, Llama,
and Gemma, respectively. The modern model overlays and evaluations also run
on the H20-3e. Peak CPU memory and production TEE latency were not
instrumented and are therefore outside the reported measurements.

\noindent\textbf{Defense diagnostics.}
For each protected tensor, we compute the sampled mean absolute output error.
Table~\ref{tab:modern_butterfly_diagnostics} reports the median and 95th
percentile of this quantity across tensors, together with the maximum sampled
error. It also reports correspondence recovery before and after removing 1,
8, or 32 exposed principal components.

\section{Additional Scope and Limitations}
\label{sec:appendix_discussion}

\noindent\textbf{Relation to LWE.}
Neither \attackname nor \blockhlpname solves Learning with Errors
(LWE)~\cite{regev2005lwe}. The original analysis derives a relation among the
public ancestor, victim weights, and exposed weights, quantizes it over a
large prime field, and interprets the resulting expression through an
LWE-style argument. The released implementation, however, performs real
arithmetic and does not instantiate standard LWE.

To evaluate the stronger modular security intuition directly, we construct a
defense-favorable realization in which the exposed weights and secret
coefficients participate in arithmetic over the complete field. This
realization retains the protocol's structured diagonal-plus-rank-one
transformation. Our result is therefore narrower than an attack on LWE: it
shows that the concrete structured matrices available to the accelerator do
not obtain confidentiality merely from an analogy to LWE.

\noindent\textbf{Scope of the maximal-rank mitigation.}
\defensename removes the shared rank-one primitive instead of increasing its
rank by a small constant. In our experiments, correspondence and adaptive
PCA recovery fall to near-random levels without reducing BERT accuracy and
remain below $0.21\%$ on the three larger language models. The transferred
\blockhlpname attack recovers no victim row in the full-BERT stress test.

These results establish resistance to the attacks evaluated in this work,
not cryptographic indistinguishability. The butterfly key family remains
structured, and we do not rule out an adaptive attacker that jointly
estimates public weights, diagonal scales, permutations, and butterfly
parameters. The mask strength also introduces a numerical tradeoff:
increasing it improves directional hiding but increases floating-point
cancellation during trusted correction.

Our implementation is a correctness-oriented prototype rather than a
production TEE benchmark. Although its arithmetic count remains below dense
matrix multiplication, future work should evaluate optimized trusted
implementations, define a matrix-level security game, analyze adaptive
factorization attacks, and compare measured trusted cost with cryptographic
and TEE-only baselines under the same confidentiality target.

\noindent\textbf{Conditions for exact lattice recovery.}
The exact \blockhlpname result depends on fixed-point victim combinations
remaining sufficiently small relative to $Q$. Exact encoded row recovery also
requires the relevant row spans to be independent and each recovered integer
row to have a known scale. In our implementation, dividing by the coordinate
gcd returns the primitive representative. If the encoded victim row is not
primitive, the absolute integer scale remains ambiguous without an additional
source of scale information.

We audit these conditions in our exact-recovery evaluation. Every protected nonzero row is primitive in the evaluated BERT-Base and GPT2-Base checkpoints, and the required short-combination condition holds across all reported configurations and independent keys. Other weight distributions and quantization protocols require separate measurement.

\noindent\textbf{Finite field realization.}
The finite field (mod-Q) version follows the reduction to LWE of ArrowCloak.
The original system does not specify a complete layer by layer finite field
activation protocol covering signed rescaling, rounding, and wrap semantics.
Our modular realization therefore evaluates exact protection and recovery of
encoded weights, followed by ordinary floating point model inference. It
does not claim to implement an unspecified end-to-end integer inference
system. This construction deliberately favors the defense by applying
field-wide randomness and modular arithmetic consistently to the exposed
weights and secret coefficients.

\noindent\textbf{Role of public ancestry.}
Public ancestry serves different purposes in the two attacks. \attackname
uses the public ancestor to recover vector correspondences and to complete
the one-dimensional weight component removed during projection.
\blockhlpname instead recovers primitive integer row representatives
algebraically and uses the public weights only to determine their signs and
semantic positions.

Without a public ancestor, \blockhlpname still returns the complete unordered
projective row set under the stated recovery conditions, but converting that
set into an executable model requires another source of row labels. This
limitation narrows the final model reconstruction claim without changing the
demonstrated breach in the public ancestry deployment setting considered by
\sysname.

\end{document}